\documentclass[preprint1]{aastex}

\slugcomment{Accepted for Publication in \it The Astronomical Journal\rm}

\shorttitle{Gibson et~al.}
\shortauthors{HVC Complex~C: Galactic Fuel or Galactic Waste?}

\begin{document}
\baselineskip=14pt

\title{High-Velocity Cloud Complex~C: Galactic Fuel or Galactic Waste?}

\author{Brad K. Gibson\altaffilmark{1,2},
	Mark L. Giroux\altaffilmark{2,3},
	Steven V. Penton\altaffilmark{2}, \break
	John T. Stocke\altaffilmark{2}, 
	J. Michael Shull\altaffilmark{2,4} and
        Jason Tumlinson\altaffilmark{2} }
\altaffiltext{1}{Centre for Astrophysics Supercomputing,
		 Swinburne University,
	         Mail \#31, P.O. Box 218,
	         Hawthorn, Victoria, 3122
		 Australia}
\altaffiltext{2}{Center for Astrophysics \& Space Astronomy,
		 Department of Astrophysical \& Planetary Sciences, 
	         Campus Box 389, University of Colorado,
	         Boulder, CO 80309-0389}
\altaffiltext{3}{Department of Physics \& Astronomy, East
                 Tennessee State University, Johnson City, TN}
\altaffiltext{4}{Also at JILA, University of Colorado and National
		 Institute of Standards and Technology}

\def\spose#1{\hbox to 0pt{#1\hss}}
\def\simlt{\mathrel{\spose{\lower 3pt\hbox{$\mathchar"218$}}
     \raise 2.0pt\hbox{$\mathchar"13C$}}}
\def\simgt{\mathrel{\spose{\lower 3pt\hbox{$\mathchar"218$}}
     \raise 2.0pt\hbox{$\mathchar"13E$}}}
\def\kms{{\rm km\,s$^{-1}$}}
\def\eg{{\rm e.g.}}
\def\ie{{\rm i.e.}}
\def\etal{{\rm et~al.}}

\begin{abstract}
\baselineskip=14pt
We present HST GHRS and STIS observations of five QSOs that probe the 
prominent high-velocity cloud (HVC) Complex~C, covering $\sim$10\% of 
the northern sky.  Based upon a single sightline measurement (Mrk~290), 
a metallicity [S/H]=$-$1.05$\pm$0.12 has been associated with Complex~C
by Wakker \etal\ (1999a,b).  When coupled with its inferred distance 
($5\simlt d\simlt 30$\,kpc) and line-of-sight velocity 
($v\sim -100$ to $-$200~km\,s$^{-1}$), Complex~C appeared to
represent the first direct evidence for infalling low-metallicity
gas onto the Milky Way, which could provide the bulk of the fuel for 
star formation in the Galaxy.  We have extended the abundance analysis 
of Complex~C to encompass five sightlines.  We detect \ion{S}{2} 
absorption in three targets (Mrk~290, Mrk~817, and Mrk~279); the 
resulting [\ion{S}{2}/\ion{H}{1}] values range from $-$0.36 (Mrk~279) 
to $-$0.48 (Mrk~817) to $-$1.10 (Mrk~290).  Our preliminary \ion{O}{1}
FUSE analysis of the Mrk~817 sightline also supports the conclusion
that metallicities as high as 0.3$\times$ solar are encountered within
Complex~C.  These results complicate an 
interpretation of Complex~C as infalling low-metallicity 
Galactic fuel.  Ionization corrections for \ion{H}{2} and \ion{S}{3} 
cannot easily reconcile the higher apparent metallicities along the 
Mrk~817 and Mrk~279 sightlines with that seen toward Mrk~290, since  
H$\alpha$ emission measures preclude the existence of sufficient 
\ion{H}{2}.  If gas along the other lines of sight has a similar 
pressure and temperature to that sampled toward Mrk~290, the predicted 
H$\alpha$ emission measures would be $\sim$900\,mR.  It may be 
necessary to reclassify Complex~C as mildly enriched Galactic 
{\it waste} from the Milky Way or processed gas torn from
a disrupted neighboring dwarf, as opposed to low-metallicity Galactic 
{\it fuel}. 
\end{abstract}


\section{Introduction}
\label{intro}

The infall of low-metallicity gas onto the Milky~Way is a crucial component
of virtually all Galactic formation and evolution scenarios.  The
simple closed-box model of chemical evolution 
assumes no infall, inflow, or 
outflow within the given region being modeled,
and is successful at explaining
many of the observational characteristics of the Galactic disk. 
However, it fails 
to reproduce the metallicity distribution of nearby G- and K-dwarfs, the
so-called ``G-dwarf problem'' (Pagel 1994; Flynn \& Morell 1997),
related to the well-known paucity of metal-poor G- and K-dwarfs
in the solar neighborhood.

Closed-box models, in which
star formation proceeds as some functional form dependent upon local
gas density, tend to overproduce low-metallicity stars before self-enrichment
affects subsequent generations of stars.
Conversely, infall models are attractive because Galactic disk formation
proceeds by gradual accretion of pristine or partially processed material.
The initial reservoir of $Z=0$ gas in a given volume of the disk is presumed to
be negligible, and the number of first generation stars is likewise 
negligible.  The processed ejecta of this first generation mixes
with infalling low-metallicity gas, and a second, more substantial,
generation of star formation begins, now with $Z>0$.
This process continues, presumably unabated, to the present day,
regulated in a manner that recovers the local G-dwarf metallicity
distribution and avoids the overproduction of metal-poor disk 
stars (\eg, Pagel 1994; Figure~23).  

It should be stressed that infalling fuel does not necessarily need to
be pristine in order to satisfy the G-dwarf problem.  Under the 
scenario explored by Tosi (1988), for example, infalling gas
metallicities\footnote{Under the assumption of scaled-solar abundance
ratios, [O/H] can be replaced by [S/H] for the arguments that follow.}
[O/H]$\simlt$$-$1.0 yield results essentially indistinguishable from 
primordial metallicity gas. 
Infalling gas metallicities [O/H]$\simgt$$-$0.5 can be ruled out, owing to 
their inability to recover the present-day radial distribution of
\ion{H}{2} region oxygen abundances in the Galactic disk;
intermediate-metallicities $-$1.0$\simlt$[O/H]$\simlt$$-$0.5 are 
marginally consistent with this same \ion{H}{2} region constraint.
More sophisticated models, based upon the dual infall scenario of 
Chiappini \etal\ (1999), are currently being examined and are
discussed elsewhere (Gibson et~al. 2001,2002).  

While tenable theoretically, observational proof of the existence of this
infalling low-metallicity Galactic fuel had been difficult to obtain
prior to the work of Wakker \etal\ (1999a,b).  Larson (1972) suggested 
that the population of Galactic High-Velocity Clouds (HVCs) was the most 
likely culprit, a scenario that has been explored in detail by Tosi (1988).  
HVCs are defined as neutral hydrogen clouds travelling at
velocities incompatible with those allowed by differential Galactic
rotation (\eg, Wakker, van Woerden \& Gibson 1999c, and references therein).
Despite the ubiquity of HVCs, with a sky covering factor as great as 37\%
(Murphy, Lockman \& Savage 1995), our ignorance concerning their distance 
and metallicity makes an unambiguous identification of the Galactic HVCs 
as infalling Galactic fuel very difficult.

Wakker \etal\ (1999a,b) derive a metallicity 
of [S/H]=$-$1.05$\pm$0.12 for one portion of HVC Complex~C,
using the background Seyfert Mrk~290 as a probe of the intervening gas.
This metallicity is $3-5$ times lower, for example, than that encountered in
the Magellanic Stream ([\ion{S}{2}/\ion{H}{1}]=$-$0.5; Gibson \etal\ 2000). 
This metallicity is also consistent with Tosi's infalling Galactic fuel model.  
Wakker \etal\ conclude that Complex~C is a proto-typical
example of infalling low-metallicity gas responsible for fueling
Galactic star formation.  As they discuss, the \it source\rm~of
this infalling gas could be primordial gas left over from the
formation of the Galaxy, or from mildly-processed gas stripped
from local group dwarfs, and remains undetermined.  More metal-rich HVCs
must have another association, such as tidal
disruption of the Magellanic Clouds (Lu \etal\ 1998; Gibson \etal\ 2000) or
Galactic fountains (Richter \etal\ 1999) (Galactic waste) and
are unlikely candidates for Galactic fuel.  Because of the implications
for Galactic formation and evolution, we have undertaken further analysis 
of Complex~C to confirm the inferred low metallicity through studies of
four additional sightlines.

In \S~\ref{data}, we present the data and analysis for
five HST GHRS and STIS targets, Mrk~290, 817, 279, 501, and 876. 
Each target probes the intervening gas associated with HVC Complex~C.
Because sulfur is little affected by depletion onto dust
(Savage \& Sembach 1996; Figure~6), the column
densities provided by the \ion{S}{2} $\lambda\lambda$1250,1253,1259\, 
lines yield a robust measure of the metallicity of Complex~C and form 
the basis for much of our analysis.  The HVC gas probed by Mrk~290
appears to be of low metallicity ($\sim$0.1$\times$ solar), confirming the
work of Wakker \etal\ (1999a,b).  The Mrk~817 and Mrk~279
sightlines, however, probe Complex~C gas that appears to be more 
metal-rich ($\sim$0.2$-$0.4$\times$ solar).  Conversely, the neutral 
nitrogen abundance along the Mrk~876 sightline is $<$0.1$\times$ solar.
We attempt to reconcile these new observations with the Mrk~290 data
by invoking variable ionization conditions within Complex~C.  To first 
order, such a reconciliation is difficult.  We summarize our results in 
\S~\ref{summary} and briefly discuss the implications of
known published and preliminary metallicity measurements
of Complex~C made with FUSE.  We then suggest future observational tests 
with HST, FUSE, and 21\,cm telescopes that may 
unequivocally discriminate between the low-metallicity Galactic fuel
and associations appropriate for higher metallicity gas.

\section{Analysis}
\label{data}

The analysis of Wakker \etal\ (1999a,b) was restricted to a
single probe of Complex~C, provided by the background Seyfert Mrk~290.
A byproduct of the Stocke-Shull HST GHRS and STIS program 
on the origin and physical
conditions in the low-redshift Ly$\alpha$ forest 
(HST PID\#6593$+$7345) has been the
serendipitous detection of several major high- and intermediate-velocity 
cloud complexes (Penton, Stocke \& Shull 2000).
Our analysis of Magellanic Stream absorption features
is described elsewhere (Gibson \etal\ 2000).

Five of the 31 targets in our full GHRS$+$STIS sample, including
Mrk~290, intersect the foreground HVC Complex~C.
Figure~\ref{fig:map} shows a 2100\,deg$^2$ region of Complex~C,
constructed from the Leiden-Dwingeloo \ion{H}{1} Survey (Hartmann \&
Burton 1997; hereafter, LDS).  To avoid confusion with Galactic low- and
intermediate-velocity gas, we show only the velocity range 
$-250<v_{\rm LSR}<-110$\,km\,s$^{-1}$ in 
Figure~\ref{fig:map}.  The five targets discussed in the present
Complex~C abundance analysis (Mrk~290, 817, 279, 501, and 876)
are labeled appropriately. 

\placefigure{fig:map}

The Complex~C \ion{H}{1} column densities N(\ion{H}{1})
along each sightline span an order of magnitude, ranging from 
$1.6\times 10^{19}$\,cm$^{-2}$ (Mrk~876 sightline) to
$11.5\times 10^{19}$\,cm$^{-2}$ (Mrk~290).  We use the Effelsberg
\ion{H}{1} data of Wakker \etal\ (2001) in the analysis that follows;
the relevant column densities are listed in column 6
of Table~\ref{tbl:summary}.
It is important to bear in mind that these \ion{H}{1} column densities
are derived using the 10$^\prime$ Effelsberg beam, while the
absorption measurements discussed below sample gas
at far higher sub-arcsecond scale resolution.  It
remains an open question as to how accurately \ion{H}{1}
emission radio measurements represent the gas that
is sampled in the pencil beam represented by absorption
measurements.  Variations as large as a factor of five in N(\ion{H}{1})
may be encountered at the arcminute level (Wakker \& Schwarz 1991).  
However, for
seven targets for which both $<$1$^\prime$ and 10$^\prime$$-$12$^\prime$
\ion{H}{1} data exists (Table~1 of Wakker \etal\ 2001), the ratio of
N(\ion{H}{1})$_{1^\prime}$/N(\ion{H}{1})$_{10^\prime}$ is only
0.90$\pm$0.13 (with extrema of 0.75 and 1.24).  This might lead one to conclude
that such spatial resolution limitations are not as critical as first imagined,
although this conclusion is based upon only seven data points.

Comparisons have also been made of N(\ion{H}{1}) derived from
21~cm mapping with measurements of N(\ion{H}{1}) derived
from Ly$\alpha$ absorption along the lines of sight
to early type stars in the Galactic halo
(Lockman, Hobbs \& Shull 1986)
and toward extragalactic sources (Savage \etal\ 2000).
The ratio N(\ion{H}{1})$_{\rm Ly\alpha}$ / N(\ion{H}{1})$_{\rm 21cm}$
is somewhat less than unity, with a variation 
of less than $\pm$50\%, in agreement with the
N(\ion{H}{1})$_{1^\prime}$/N(\ion{H}{1})$_{10^\prime}$ comparison noted above.

In summary, we adopt a conservative
first-order estimate of the uncertainty of the adopted
N(\ion{H}{1}) for each sightline
due to this mismatch in resolution
of $\pm$0.2\,dex ($\pm$50\%).
It remains possible 
that we may have been unfortunate enough to encounter a line of sight
with a 10$^\prime$-inferred N(\ion{H}{1}) that differs from the
``pencil beam'' N(\ion{H}{1})
by a factor of five or more.  Higher resolution radio mapping will
better address this possibility.

Table~\ref{tbl:obs} lists the 12 HST spectra, distributed over the
five targets employed in the present study; all but one (Mrk~876)
are GHRS G160M data.  Of the GHRS data, only one spectrum (Mrk~501)
was a pre-COSTAR observation.  The data for Mrk~290 (PI: Wakker; 
PID \#6590) and Mrk~876
(PI Cot\'e: PID \#7295) were extracted from the HST Archive, to
supplement the three targets from our GO programs.  Details concerning
spectrum preparation are described
by Penton \etal\ (2000).

\placetable{tbl:obs}

In deriving the metallicity of Complex~C, 
we have employed two different techniques,
one based upon the assumption of optically-thin absorption
features, and the other based upon the apparent optical depth method of Sembach
\& Savage (1992).
In the weak-line (optically-thin) limit, 
the column density N$_{\tau=0}$ (in cm$^{-2}$)
and equivalent width of a line
W$_\lambda$ (in m\AA) are related through
\begin{equation}
{\rm N}_{\tau=0} = 1.13\times 10^{17}\,{{{\rm W}_\lambda}\over{f\lambda_0^2}}, 
\label{eq:column}
\end{equation}
\noindent
where $\lambda_0$ (in \AA) is the rest wavelength of the line and $f$ is its
oscillator strength (\eg, Savage \& Sembach 1996; equation 3).  For the lines
considered in this paper, the relevant values of $\lambda_0$ and $f$
are provided in columns 2 and 3 of Table \ref{tbl:atomic}.

\placetable{tbl:atomic}

Under the apparent optical depth method of Sembach \& Savage (1992), the
apparent column density N$_{\tau_v}$, in the limit of a finite number of data
points, is given by
\begin{equation}
{\rm N}_{\tau_v} = {{3.767\times 10^{14}}\over{f\lambda_0}}\,\sum_{i=1}^n
\ln\biggl[{{I_c(v_i)}\over{I(v_i)}}\biggr]\,{\rm d}v_i,
\label{eq:app_col}
\end{equation}
\noindent
where $I(v_i)$ and $I_c(v_i)$ are the observed and estimated continuum
intensities at velocity $v_i$ (equations A21 and A29 of Sembach \& Savage).
The statistical noise uncertainty associated with the apparent column density
of equation~\ref{eq:app_col} is
\begin{equation}
\sigma_{{\rm N}_{\tau_v}} = {\rm N}_{\tau_v}\,\Biggl(
\sqrt{\sum_{i=1}^n\sigma^2(v_i)\big/I(v_i)^2\,({\rm d}v_i)^2}
\Biggr/\sum_{i=1}^n\ln\biggl[{{I_c(v_i)}\over{I(v_i)}}\biggr]\,{\rm d}v_i
\Biggr),
\label{eq:sig_app_col}
\end{equation}
\noindent
based upon equations~A27 and A30 of Sembach \& Savage.  Uncertainties quoted in
this analysis correspond to those of equation~\ref{eq:sig_app_col}; those
associated with continuum placement have not been considered here.  For the
high S/N GHRS data, the uncertainty associated with the
continuum placement is only an additional $\sim$5\% beyond that of
equation~\ref{eq:sig_app_col} (Penton \etal\ 2000) and is
neglected here.

Generally, our Complex~C metallicity determinations are made via the
singly-ionized \ion{S}{2}\,$\lambda\lambda$1250,1253,1259
lines.  \ion{S}{2} is a good species for these measurements because:
(i) sulfur is only mildly depleted onto dust, and (ii) \ion{S}{2}
is the dominant ionization stage in the cold and warm photoionized portions of
the interstellar medium (Savage \& Sembach 1996).  
It remains true that we do not directly measure [{\rm S/H}], the
sulfur abundance with respect to that of the Sun, but rather S~II/H~I.  
Our assumed solar abundances for relevant elements may be found in 
Table \ref{tbl:atomic}.

In \S\S~\ref{mrk290}-\ref{mrk876}, we present our GHRS$+$STIS
spectra for each of the five background
probes of HVC Complex~C.  We provide detections and 
upper limits for N(\ion{S}{2}), for Mrk~290, 817, and 279, which,
in combination with the \ion{H}{1} column densities of 
Table~\ref{tbl:summary}, these yield
the ratio \ion{S}{2}/\ion{H}{1}.  
Uncertainties in the conversion of this ratio to the \it true \rm
metallicity [S/H] will be discussed for the individual sightlines.  

\subsection{Mrk~290}
\label{mrk290}

Wakker \etal\ (1999a,b) were the first to derive a metallicity for HVC
Complex~C.  Using the \ion{S}{2}\,$\lambda$1253 and 
\ion{S}{2}\,$\lambda$1259 lines, they found 
[S/H]$\equiv$[(\ion{S}{2}$+$\ion{S}{3})/(\ion{H}{1}$+$\ion{H}{2})]=
$-$1.05$\pm$0.12, under the assumption that, within the portion
of the cloud where hydrogen is fully ionized, sulfur is
2/3 \ion{S}{2} and 1/3 \ion{S}{3},
and N(\ion{H}{2})=0.2\,N(\ion{H}{1}).  The latter was derived empirically
from their WHAM H$\alpha$ emission measure analysis of the Mrk~290 sightline,
for which I(H$\alpha$)=187$\pm$10\,mR.  The former assumption is
more uncertain, as only upper limits on \ion{S}{2}\,$\lambda$6716
emission
were available for this field.  However, with such a small inferred
\ion{H}{2} fraction, their conclusions were insensitive to
the assumed \ion{S}{3}/\ion{S}{2} ratio in the \ion{H}{2} region.

Figure~\ref{fig:mrk290_ghrs} shows a
10\,\AA\ region of the
co-added GHRS G160M spectrum used in our analysis.  This spectrum has
been smoothed by the post-COSTAR line spread
function (Gilliland 1994) to improve the signal-to-noise (S/N), 
although we measure
absorption feature equivalent widths from the raw spectra.  
The Galactic (G) and Complex~C (C) 
\ion{S}{2} $\lambda$1253\, and \ion{S}{2} $\lambda$1259\, lines are clearly
seen, as is the saturated blend of (G$+$C) \ion{Si}{2}\,$\lambda$1260.
The latter, because of this saturation,
provides only a loose lower limit on the silicon abundance along
this sightline.  The \ion{S}{2}\,$\lambda$1253\,
line resides on the redward wing of the broad, intrinsic Mrk~290 Ly$\alpha$
emission line, complicating continuum placement.  

\placefigure{fig:mrk290_ghrs}

Normalizing by the local continua, we obtain the 
\ion{S}{2}\,$\lambda$1259, \ion{S}{2}\,$\lambda$1253, and
\ion{Si}{2}\,$\lambda$1260 profiles shown in
Figure~\ref{fig:mrk290_stack}.  The abscissa has been transformed from 
the GHRS heliocentric wavelength scale to velocity with respect to
the local standard of rest (LSR).  
An \it a posteriori \rm shift of 
$+$20\,km\,s$^{-1}$ was applied, in order to reconcile a residual systematic
offset between the Galactic and HVC \ion{S}{2} absorption features
(middle two panels) and \ion{H}{1} emission features (upper panel).
The GHRS line profiles shown in Figure~\ref{fig:mrk290_stack} are the raw
data, unsmoothed by the post-COSTAR GHRS line spread
function.
The Effelsberg \ion{H}{1} spectrum for this sightline is shown in the upper 
panel of Figure~\ref{fig:mrk290_stack}, and was kindly provided by Peter
Kalberla and Bart Wakker prior to publication (Wakker \etal\ 2001).

Two individual Complex~C components are clearly 
seen\footnote{Wakker \etal\ (1999a,b) state that the $v_{\rm
LSR}\approx -105$\,km\,s$^{-1}$ component of \ion{S}{2} $\lambda$1259\, 
is missing in the GHRS spectrum.  We believe we have detected it, based 
upon the middle panel of Figure~\ref{fig:mrk290_stack}.}
in both \ion{H}{1} and \ion{S}{2} $\lambda$1259\, but only marginally so 
in \ion{S}{2} $\lambda$1253.  The \ion{S}{2} and \ion{H}{1} centroids 
for both components agree to within 4\,km\,s$^{-1}$.  We have not 
deconvolved the line profiles in order to treat components 1 and 2 
separately; instead, we have simply integrated over the full line profile 
in both the Effelsberg and HST datasets, just as we did in our earlier 
analysis of the Magellanic Stream (Gibson \etal\ 2000).

\placefigure{fig:mrk290_stack}

In Table~\ref{tbl:features}, the first three entries 
list the relevant information for the
\ion{S}{2} and \ion{Si}{2} detections of Complex~C in the Mrk~290
sightline.  The line centroid (column 3) and velocity range over which the line
profile was integrated (column 4) yield the quoted
equivalent width (column 5).  These
are supplemented by the previously mentioned two
determinations of the inferred column density, N$_{\tau=0}$ (optically-thin
assumption - column 6) and N$_{\tau_v}$ (apparent optical depth method of
Sembach \& Savage 1992 - column 7).  Our \ion{S}{2}\,$\lambda\lambda$1253,1259
equivalent widths are
indistinguishable from those of Wakker \etal\ (1999a,b), and consistent with
the expected theoretical ratio (1.5), suggesting that optical depth effects
are minimal.

\placetable{tbl:features}

Using equations~\ref{eq:app_col} and \ref{eq:sig_app_col}, the apparent column
densities of the Complex~C \ion{S}{2} and \ion{Si}{2} features seen in the
Mrk~290 GHRS spectrum are
N$_{\tau_v}$(\ion{S}{2}\,$\lambda$1259)=
(1.81$\pm$0.24)$\times$10$^{14}$\,cm$^{-2}$,
N$_{\tau_v}$(\ion{S}{2}\,$\lambda$1253)=
(1.65$\pm$0.17)$\times$10$^{14}$\,cm$^{-2}$, and
N$_{\tau_v}$(\ion{Si}{2})$>$1.0$\times$10$^{14}$\,cm$^{-2}$.
The \ion{S}{2}-weighted average is N$_{\tau_v}$(\ion{S}{2})$\equiv$
N(\ion{S}{2})=(1.70$\pm$0.14)$\times$10$^{14}$\,cm$^{-2}$.
With a solar abundance A$_{\rm S_\odot}=(1.862\pm 0.215)\times 10^{-5}$ 
(column 4 of Table~\ref{tbl:atomic}) and 
N(\ion{H}{1})=(11.5$\pm$0.5)$\times$10$^{19}$\,cm$^{-2}$
(Table~\ref{tbl:summary}), we can now write the \ion{S}{2} and \ion{Si}{2}
abundances for Complex~C (sampled by Mrk~290), as
\begin{eqnarray}
{[\rm S~II/H~I]}  & = & -1.10\pm 0.06 \nonumber \\
{[\rm Si~II/H~I]} & > & -1.61\pm 0.03 \nonumber
\end{eqnarray}
\noindent
The above quoted uncertainty only represents the random statistical component;
a conservative systematic uncertainty of $\pm$0.10\,dex is assumed for both 
N(\ion{H}{1}) (Wakker 2001) and the GHRS continuum placement (Penton \etal\
2000), for a total systematic uncertainty of $\pm$0.14\,dex.

Of potentially greater concern are the ionization corrections required to 
derive [S/H] from our [\ion{S}{2}/\ion{H}{1}] determination.
Based on their WHAM data, Wakker \etal\ (1999a,b) find
N(\ion{H}{2})=(0.20$\pm$0.07)\,N(\ion{H}{1}), and
under the assumption that
within the portion of the cloud where hydrogen is fully ionized, 
0$-$100\% of the sulfur is
in the form of \ion{S}{3}, 
N(\ion{S}{3})=(0$-$0.2$\pm$0.07)\,N(\ion{S}{2}).
This implies a correction that may reach
0.08$\pm$0.03\,dex, if N(\ion{S}{3}) = 0,
in the sense of 
[S/H]=[\ion{S}{2}/\ion{H}{1}]$-$0.08. 
However, one must bear in mind that the
WHAM field of view is 1$^\circ$, and the same resolution uncertainties
that plague
\ion{H}{1} column density determinations 
also affect the \ion{H}{2} measurements.  
In addition, for systems with larger fractions of \ion{H}{2},
uncertainties in the fraction of \ion{S}{3} present in these
ionized regions become increasingly important.
Because we have, as yet, no emission measure information for any of our 
remaining sightlines,\footnote{Murphy \etal\ (2000) recently
reported the non-detection of H$\alpha$ in the WHAM data for the Mrk~876
sightline; we return to this in \S~\ref{mrk876}.}
we generally restrict our
discussion to [\ion{S}{2}/\ion{H}{1}], but draw attention
to important implications for future emission measure work, where appropriate.

As an additional way to explore the uncertainties associated with
converting [\ion{S}{2}/\ion{H}{1}] to [S/H], we have constructed
a grid of photoionization models using the code Cloudy (Ferland 1996).
We make the very simplified assumption that the absorbers are
plane-parallel slabs of gas illuminated on one side by incident
ionizing radiation dominated by Galactic OB associations.  We 
approximate this integrated spectrum with a power-law $\nu^{-2}$
spectrum between 1 and 4 Rydbergs, with a factor of 100 drop
beyond 4 Rydbergs, which includes the harder extragalactic
component.  We assume a level of ionizing radiation compatible
with Complex~C being at a distance of order 10$-$20\,kpc, and
exposed to approximately 5\% of the ionizing radiation produced
in the Galaxy, that is, a one-sided, normally incident photon flux 
(photons cm$^{-2}$ s$^{-1}$) of $\log\phi\approx 5.5$.  While this
power law is a reasonable approximation of the integrated spectrum of 
OB associations (Sutherland \& Shull 2001), we also consider
a softer spectrum, that based upon a T=35,000\,K Kurucz model
atmosphere.  Such a soft ionizing spectrum has been
suggested to account for emission-line ratios
in the warm interstellar medium (Sembach \etal\ 2000).

\placefigure{fig:SII}

Figure~\ref{fig:SII} summarizes the results of our calculations.
The top panel, which assumes the absorber has 
N(\ion{H}{1})=10$^{20}$\,cm$^{-2}$, indicates that little ionization
corrections are required for high column density
lines of sight such as those toward Fairall~9 (Gibson \etal\ 2000).  
Even for the case when the density $n_{\rm H} = 0.01$\,cm$^{-3}$ and 
metallicities are 0.1 of solar, [\ion{S}{2}/\ion{H}{1}] provides only
a 50\% overestimate of [S/H].  For higher densities,
the correction is negligible.  For much lower
column densities N(\ion{H}{1})=10$^{19}$\,cm$^{-2}$, 
the bottom panel shows that the ionization correction remains
small for $n_{\rm H}\simgt 0.1$\,cm$^{-3}$.

\subsection{Mrk~817}
\label{mrk817}

Figure~\ref{fig:mrk817_ghrs} shows a 16\,\AA\ region of our GHRS G160M co-added
spectrum of Mrk~817, centered on the intervening Galactic
\ion{S}{2}\,$\lambda$1250 feature.  
The spectrum was smoothed with
the post-COSTAR line spread function (Gilliland 1994).  As was the case for
Mrk~290, Galactic (G) and Complex~C (C) \ion{S}{2}
features, here at 1250\,\AA\ and 1253\,\AA, are clearly seen in absorption.
Because the \ion{S}{2}\,$\lambda$1253 features 
coincide with the peak of the intrinsic
Ly$\alpha$ emission line
associated with Mrk~817, identification of the local continuum 
is more challenging than it was for Mrk~290, although we retain confidence
in the third-order polynomial employed.

\placefigure{fig:mrk817_ghrs}

The normalized 
velocity stack of relevant \ion{S}{2} absorption features is shown
in Figure~\ref{fig:mrk817_stack}.  
We employed a shift of
$+$20\,km\,s$^{-1}$ to bring the centroids of the Galactic
\ion{S}{2} features into alignment with those of 
the Galactic \ion{H}{1} (upper panel of Figure~\ref{fig:mrk817_stack}).
The Complex~C \ion{H}{1} column density along this sightline is
N(\ion{H}{1})=(3.0$\pm$0.1)$\times$10$^{19}$\,cm$^{-2}$ (Wakker \etal\ 2001).

\placefigure{fig:mrk817_stack}

The equivalent widths for the two Complex~C \ion{S}{2} absorption features, 
W$_\lambda$(1253\,\AA)=27$\pm$3\,m\AA\ and
W$_\lambda$(1250\,\AA)=13$\pm$3\,m\AA\ (see relevant entries in
Table~\ref{tbl:features}), are consistent with the expected
theoretical ratio (2.0), again suggesting that saturation effects are 
minimal, as they were for the Mrk~290 sightline.
Using equations~\ref{eq:app_col} and \ref{eq:sig_app_col},
we find the \ion{S}{2} column density to be
N(\ion{S}{2})=$(1.89\pm 0.13)\times 10^{14}$\,cm$^{-2}$.

Coupling the \ion{S}{2} and \ion{H}{1} column densities, we 
infer the Complex~C sulfur abundance along this
Mrk~817 sightline to be
\begin{eqnarray}
{[\rm S~II/H~I]} & = & -0.48\pm 0.06, \nonumber
\end{eqnarray}
\noindent
a factor of 4 times greater than the inferred [\ion{S}{2}/\ion{H}{1}]
along the line-of-sight to Mrk~290 (\S~\ref{mrk290}).  Again,
the quoted error budget reflects internal statistical uncertainties alone.

It is tempting to now explore whether our Mrk~817 result, 
[\ion{S}{2}/\ion{H}{1}]=$-$0.48$\pm$0.06, could be reconciled
with the Wakker \etal\ (1999a,b) result for Mrk~290, [S/H]=$-$1.05$\pm$0.12, 
via ionization corrections.  We assume a simple model where the
gas is either completely neutral in H or completely ionized in H.
In the \ion{H}{1} region, S is entirely \ion{S}{2}, being photoionized by
diffuse ultraviolet radiation.  In the
\ion{H}{2} region, S is divided between \ion{S}{2} and \ion{S}{3},
with a ratio $r=$\ion{S}{3}/\ion{S}{2}.  In this case, it may
be shown that the additional ionized hydrogen column density
N(\ion{H}{2}) which must be invoked to reconcile
[\ion{S}{2}/\ion{H}{1}] with [S/H] is
\begin{equation}
{ {\rm N(H~II) } \over {\rm N(H~I)} } =
\left( 1+r \right)
\left[ 10^{\rm [SII/HI] - [S/H]} - 1 \right].
\label{eq:HIIcorrection}
\end{equation}
\noindent
Under the assumption that the Mrk~290
metallicity determination ([S/H]=$-$1.05) is correct, and
$r=0.5$, 
equation~\ref{eq:HIIcorrection} implies that a column density
N(\ion{H}{2})=4.1\,N(\ion{H}{1}) must be present to be 
consistent with [\ion{S}{2}/\ion{H}{1}].\footnote{For $r$=0 (1),
N(\ion{H}{2})=2.7 (5.4) N(\ion{H}{1}).}

This estimate represents an unreasonably large amount of additional 
ionized hydrogen, for reasons that we now explain.
Combined with the Effelsberg \ion{H}{1} data,
N(\ion{H}{1})$_{\rm Eff}$=$3.05\times 10^{19}$\,cm$^{-2}$,
this implies N(\ion{H}{2})$=1.24\times 10^{20}$\,cm$^{-2}$.
If we assume that the temperature and pressure of the Complex~C gas 
probed by Mrk~817 and 279 are similar to that inferred for the Mrk~290 
sightline, and adopt the Wakker \etal\ (1999a)
standard model of constant density, neutral core, and ionized shell, we can
scale the observational results provided by WHAM (see \S~\ref{mrk290}),
to yield a first-order prediction for the H$\alpha$ emission measure
I(H$\alpha)_{\rm pred}$ along the Mrk~817 sightline,
\begin{equation}
{\rm I(H\alpha)_{pred}} = {{\rm N(H~II)_{Mrk~817}}\over{\rm 
N(H~II)_{Mrk~290}}}\,{\rm I(H\alpha)_{Mrk~290}}, \label{eq:WHAM}
\end{equation}
\noindent
where N(\ion{H}{2})$_{\rm Mrk~290}$=$(1.93\pm 0.52)\times 10^{19}$\,cm$^{-2}$ 
and I(H$\alpha)_{\rm Mrk~290}$=187$\pm$10\,mR.  For 
N(\ion{H}{2})$_{\rm Mrk~817}$=$1.24\times 10^{20}$\,cm$^{-2}$, 
equation~\ref{eq:WHAM} predicts an H$\alpha$
emission measure of I(H$\alpha)_{\rm pred}$$\approx$1200\,mR.
Arbitrarily setting N(\ion{S}{3})=0 (\ie, not allowing
\ion{S}{3} to track \ion{H}{2}, contrary to expectations) for Mrk~817,
the inferred N(\ion{H}{2}) would be reduced to 2.7\,N(\ion{H}{1}).
The predicted H$\alpha$ emission measure, under this scenario, is
I(H$\alpha)_{\rm pred}$$\approx$800\,mR, still a surprisingly high result.
We emphasize, however, that this is a highly tentative suggestion
for the magnitude of I(H$\alpha)_{\rm pred}$.  If we assume a range in
$r = 0.2-0.8$ and propagate our formal uncertainties as
well as the uncertainties in the Wakker \etal\ (1999a,b)
results through equations \ref{eq:HIIcorrection}
and \ref{eq:WHAM}, this would introduce a range
in I(H$\alpha)_{\rm pred}$ $\approx$600-4000\,mR.  Inclusion
of the systematic uncertainties in our measurements would broaden
this range still further.

We estimate that the true \ion{H}{1} column density along the
Mrk~817 sightline would have to be 
N(\ion{H}{1})$_{\rm true}$$\approx$2.5\,N(\ion{H}{1})$_{\rm Eff}$, 
in order to reconcile the apparent
metallicity ([\ion{S}{2}/\ion{H}{1}]=$-$0.48) with that seen toward 
Mrk~290 ([S/H]=$-$1.05).  As we have discussed, one cannot exclude 
the suggestion that N(\ion{H}{1})$_{\rm Eff}$ underestimates the true 
\ion{H}{1} column density by a factor of 2.5.  However, despite the 
fact that N(\ion{H}{1}) toward the Mrk~290 sightline was measured 
with additional 2$^\prime$ Westerbork data, one similarly 
cannot dismiss the suggestion that its column may be overestimated by 
the same factor of 2.5, since the 2$^\prime$
Westerbork beam is still two orders of magnitude greater than the 
solid angle probed by GHRS. Indeed, the intermediate velocity gas 
in the foreground of globular clusters such as M13, M15 and M92 show 
factors of ten and more variations in N(\ion{Na}{1}).  
One also sees variations approaching a factor of 10 in
N(\ion{Ca}{2}), and possibly N(\ion{H}{1}), on 10$^{\prime\prime}$
angular scales.  At at distance $d=100$\,pc, this corresponds to
order-of-magnitude variations on 0.005\,pc spatial scales
[Shaw \etal\ 1996; Meyer \& Lauroesch 1999; Andrews, Meyer, \&
Lauroesch 2001 (see references therein for similar studies using
binary or single stars)].  
The latter scale is three
orders of magnitude smaller than the Complex~C spatial scale probed by the
2$^\prime$ Westerbork beam for $d=10$\,kpc. 
While these metal ions are generally not the dominant ion and so
may not trace \ion{H}{1} faithfully, they do
suggest that significant
\ion{H}{1} column variations at sub-arcminute scales may complicate 
both Mrk~290 and Mrk~817 analyses.
Further evidence of significant
sub-arcsecond \ion{H}{1} column density variations in the interstellar medium
are provided by Davis, Diamond, \& Goss (1996).

\subsection{Mrk~279}
\label{mrk279}

Our co-added GHRS G160M spectrum of Mrk~279 is the final sightline for which we
claim a detection of \ion{S}{2}.  Figure~\ref{fig:mrk279_ghrs} shows a
14\,\AA\ region of the spectrum employed, again centered upon the Galactic
\ion{S}{2}\,$\lambda$1250 absorption feature.  Complex~C is seen in 
\ion{S}{2}\,$\lambda$1250 absorption, as identified.  
The detection is admittedly
significant only at the 4$\sigma$ level (Penton \etal\ 2000), 
but was identified in a survey unbiased by any knowledge
of corresponding \ion{H}{1}.
It is apparent in
each of the four individual sub-exposures, and we currently see no compelling
reason to dismiss the line assignment, despite its lower significance.

\placefigure{fig:mrk279_ghrs}

The \ion{H}{1} profile
along this sightline is significantly more complex than that
encountered in any of the others included in the present study.
Wakker \etal\ (2001) decompose the full profile (upper panel of 
Figure~\ref{fig:mrk279_stack}) into eight components at
low-, intermediate-, and
high-velocity.  Two Complex~C components can be identified 
(centered at $-$102\,km\,s$^{-1}$ and $-$137\,km\,s$^{-1}$), with combined 
\ion{H}{1} column density 
N(\ion{H}{1})=(3.1$\pm$0.4)$\times$10$^{19}$\,cm$^{-2}$.
We have
made no attempt to deconvolve the \ion{S}{2}\,$\lambda$1250 
absorption feature and simply integrate both the \ion{S}{2} and \ion{H}{1}
profiles over the full velocity range spanned by the two components.
The result of the \ion{S}{2} integration is a measured
equivalent width of 18$\pm$ 3 m\AA, where the uncertainty is
a reflection of the formal statistical error.

The middle and lower panels of Figure~\ref{fig:mrk279_stack} show the
normalized GHRS spectrum in the vicinity 
of \ion{S}{2}\,$\lambda$1253\, and
\ion{S}{2}\,$\lambda$1250, respectively.  
A velocity shift of $+$20\,km\,s$^{-1}$ was applied, in
order to align the centroids of the Galactic 
\ion{S}{2}\,$\lambda$1250,1253
lines with the Galactic \ion{H}{1}.  It should be
stressed that the appropriate correction 
to the velocity scale was difficult to
ascertain.  Strong Ly$\alpha$ absorption intrinsic to
Mrk~279 is seen ``redward'' of the Galactic 
\ion{S}{2}\,$\lambda$1250
absorption, spanning $+50\simlt v_{\rm LSR}\simlt +350$\,km\,s$^{-1}$)
and complicating the velocity scale determination.
While $+$20\,km\,s$^{-1}$ was the applied shift, we estimate
that this is uncertain at the $\pm$20 km\,s$^{-1}$ level.
This additional uncertainty compounds the systematic uncertainty
arising from the assumption that the lower angular resolution
H~I map is a faithful guide to the \ion{S}{2} velocity profile.
We have explored the extent of this uncertainty by stepping our
integration from -200 to  -110 km s$^{-1}$ to
-160 to -70 km s$^{-1}$.  This can introduce
high excursions in the measured equivalent width as the integration
begins to include Galactic absorption. 
The lowest excursion in measured equivalent width is 16.1$\pm$3 m\AA.
While this is an extermal value, to be conservative
we have added this 1.9~m\AA\ excursion to the formal uncertainty.

\placefigure{fig:mrk279_stack}

Complex~C \ion{S}{2} absorption along this sightline is only apparent in the
\ion{S}{2}\,$\lambda$1250 line, at $-180\simlt v_{\rm LSR}\simlt
-90$\,km\,s$^{-1}$ (lower panel of Figure~\ref{fig:mrk279_stack}).
The characteristics of the feature (\ie, centroid, equivalent width, etc.) are
listed in Table~\ref{tbl:features}.  
The centroid of the Complex~C \ion{S}{2}\,$\lambda$1250
absorption is offset by $\sim$20\,km\,s$^{-1}$ from the \ion{H}{1}, but the
aforementioned uncertainty in the velocity scale
makes the significance of this offset unclear.  Unfortunately, the predicted
location of the expected Complex~C 
\ion{S}{2}\,$\lambda$1253 absorption coincides
with strong Ly$\alpha$ absorption intrinsic to Mrk~279 at $v_{\rm LSR}\approx
-150$\,km\,s$^{-1}$ (middle panel of Figure~\ref{fig:mrk279_stack}). Our 
conclusions here are necessarily based upon 
the single \ion{S}{2}\,$\lambda$1250 line.

With N(\ion{H}{1})=(3.1$\pm$0.4)$\times$10$^{19}$\,cm$^{-2}$ and an
apparent \ion{S}{2} column density 
N(\ion{S}{2}\,$\lambda$1250)=(2.48$\pm$0.68)$\times$10$^{14}$\,cm$^{-2}$,
equations~\ref{eq:app_col} and \ref{eq:sig_app_col} yield
\begin{eqnarray}
{[\rm S~II/H~I]} & = & -0.36\pm 0.18. \nonumber
\end{eqnarray}
\noindent
Recall that for the Mrk~290 and 817
sightlines we found [\ion{S}{2}/\ion{H}{1}]=$-$1.10$\pm$0.06 and
[\ion{S}{2}/\ion{H}{1}]=$-$0.48$\pm$0.06, respectively.  The inferred
metallicity along this Mrk~279 sightline seems surprisingly high.

As was found to be the case for Mrk~817, it seems untenable to appeal
to ionization corrections to reconcile our derived
[\ion{S}{2}/\ion{H}{1}]=$-$0.36$\pm$0.18
with the [S/H]=$-$1.05$\pm$0.12 found for Mrk~290 by Wakker \etal\ 
(1999a,b), Using equation~\ref{eq:HIIcorrection} for Mrk~279, we find
N(\ion{H}{2})=$1.79\times 10^{20}$\,cm$^{-2}$, 
assuming \ion{S}{3}/\ion{S}{2}$=$ 0.5.  
Via equation~\ref{eq:WHAM}, the predicted H$\alpha$ emission measure 
in this case would be I(H$\alpha)_{\rm pred}$$\approx$1700\,mR, 
$\sim$40\% greater than that predicted for the Mrk~817 sightline 
(\S~\ref{mrk817}), and a factor of $\sim$9 times greater than that 
seen toward Mrk~290.\footnote{And a factor of $\simgt$85 times greater 
than that seen toward Mrk~876 (Murphy \etal\ 2000).}  
The marginal detection of \ion{S}{3}\,$\lambda$1012 in our high S/N FUSE 
spectrum implies N(\ion{S}{3})$\approx$(0.3$-$0.4)$\times$10$^{14}$\,cm$^{-2}$, 
which is only 14\% that of the \ion{S}{2} column density.
This would imply N(\ion{H}{2})$\approx$1.4$\times$10$^{20}$\,cm$^{-2}$
and I(H$\alpha$)$\approx$\,1300\,mR.  Even in the limit of
N(\ion{S}{3})=0 in the ionized regions of the cloud, an \ion{H}{2}
column density of N(\ion{H}{2})$\simgt$10$^{20}$\,cm$^{-2}$ 
is needed to bring the Mrk~279 metallicity into formal 1$\sigma$ 
agreement with that inferred from Mrk~290.  In this case,
N(\ion{H}{2}) is still a factor of 5 greater than
that seen along the Mrk~290 sightline.  

Since the \ion{S}{2}\,$\lambda$1250 absorption is real and visible in
each of the four sub-exposures, we are left with two options.  Either our 
line assignment is valid and the inferred metallicity is a factor of 
$\sim$5 times higher than that along the Mrk~290 line of sight, or 
the 1250\,\AA\ feature is an intrinsic Ly$\alpha$ feature, and (perhaps) 
there is no metallicity discrepancy.  The detection of 
\ion{Fe}{2}\,$\lambda\lambda$1122,1144 and \ion{Si}{2}\,$\lambda$1020 with 
FUSE lends indirect support to our line assignment.  
Regardless, without \ion{S}{2}\,$\lambda\lambda$1253,1259,
we are unable to unequivocally rule out a line misidentification for
our claimed \ion{S}{2}\,$\lambda$1250 HVC feature.

\subsection{Mrk~501}
\label{mrk501}

For completeness, we include here our GHRS pre-COSTAR 
G160M spectrum of Mrk~501.  This background Seyfert probes a small clump of
Complex~C, approximately 30$^\circ$ from Mrk~290 (see
Figure~\ref{fig:map}).  
Figure~\ref{fig:mrk501_ghrs} shows a 16\,\AA\ region of the 
spectrum, smoothed with the pre-COSTAR line spread
function.  The identification of line-free continuum regions
in Figure~\ref{fig:mrk501_ghrs} is clearly difficult, complicated by the
significant undulations seen throughout the spectrum.  The latter can
masquerade as broad, shallow, absorption features.

\placefigure{fig:mrk501_ghrs}

The normalized spectra in the vicinity of \ion{S}{2}\,$\lambda$1253
and \ion{S}{2}\,$\lambda$1250 are shown in the middle and
lower panels, respectively, of Figure~\ref{fig:mrk501_stack}; the upper 
panel shows the Effelsberg \ion{H}{1} profile along this line of
sight.  The two-component structure of Complex~C is apparent, as it was 
for Mrk~279 (Figure~\ref{fig:mrk279_stack}).  The total \ion{H}{1} column 
density, integrating over both components, 
is N(\ion{H}{1})=(1.6$\pm$0.1)$\times$10$^{19}$\,cm$^{-2}$.

\placefigure{fig:mrk501_stack}

The continuum near \ion{S}{2}\,$\lambda$1250
is particularly poorly constrained, 
as witnessed by the middle panel of Figure~\ref{fig:mrk501_stack}, so we will
restrict the following discussion to the \it marginally \rm better-determined
continuum near \ion{S}{2}\,$\lambda$1253.  With a local continuum defined
over the $-$1400$\le v_{\rm LSR}\le +$800\,km\,s$^{-1}$ region surrounding
\ion{S}{2}\,$\lambda$1253, the non-detection of Complex~C
\ion{S}{2}\,$\lambda$1253 absorption sets an upper limit\footnote{The
local continuum employed here is to be preferred over the global one used by
Penton \etal\ (2000), at least for the analysis of Complex~C absorption.} to
N(\ion{S}{2}\,$\lambda$1253) of 2.0$\times$10$^{14}$\,cm$^{-2}$.
This conservative upper limit sets a fairly unrestrictive upper limit to the
sulfur abundance of
\begin{eqnarray}
{[\rm S~II/H~I]} & < & -0.16\pm 0.06. \nonumber
\end{eqnarray}
\noindent
Because of the uncertain local
continuum identification, we do not wish to belabor further the Mrk~501 
constraint.  Complicating the analysis still further is the fact that
there appears to be an IVC along this sightline whose
properties may have contaminated our analysis (see Wakker 2001 for details).

\subsection{Mrk~876}
\label{mrk876}

Mrk~876 was observed by Cot\'e \etal\ (PID \#7295), as part of their
STIS G140M program designed to probe the outer disk/halo of nearby
spirals. 
As Figure~\ref{fig:mrk876_stis} shows, Cot\'e
\etal\ have been successful in detecting Ly$\alpha$ absorption associated
with the halo of NGC~6140.  A second, serendipitous, detection of Ly$\alpha$
associated with a small group of galaxies at $v\approx 3500$\,km\,s$^{-1}$
is also seen.
Because the Cot\'e \etal\ program was centered on relatively low-redshift
Ly$\alpha$, their STIS G140M wavelength range did not encompass the
\ion{S}{2} lines. 
Instead, we will restrict ourselves
to the constraints imposed by the detection of Complex~C in
\ion{N}{1}\,$\lambda$1199.5 and 
\ion{Si}{3}\,$\lambda$1206.  The S/N in
this single STIS spectrum is only $\sim$8, with a 3$\sigma$ detection limit
of $\sim$58\,m\AA\ near both of these features.  Despite the low S/N, our
analysis does provide a useful complement to the FUSE Early Release 
Observations described by Murphy \etal\ (2000).  

Murphy \etal\ (2000) found [\ion{Fe}{2}/\ion{H}{1}]=$-$0.32$\pm$0.19 and
[\ion{N}{1}$+$\ion{N}{2}/\ion{H}{1}]$>$$-$0.83$\pm$0.10, for Complex~C along
the Mrk~876 line of sight.  This apparent high iron abundance is at odds with
the sulfur abundance derived toward Mrk~290, but in agreement with the higher
sulfur values derived from our Mrk~817 and 279 data.  Wakker (2001) favors a
photo-ionization interpretation to the FUSE Mrk~876 iron results, such that
N(\ion{H}{2})$\approx$3\,N(\ion{H}{1}) and N(Fe)$\approx$N(\ion{Fe}{2}), and 
therefore [Fe/H]$\approx$$-$1.0.  Such ionization ratios are unusual in the ISM.
 The non-detection of
H$\alpha$ (Murphy \etal\ 2000) is also somewhat surprising (see the related 
discussions of predicted emission measures in \S~\ref{mrk817} and
\ref{mrk279}).

\placefigure{fig:mrk876_stis}

While the overall shape of the continuum near 
\ion{N}{1}\,$\lambda$1199.5 and \ion{Si}{3}\,$\lambda$1206
is well-defined, the low S/N makes setting its ``zero point'' somewhat
uncertain.  Regardless, normalizing by the best-fit third-order polynomial
leads to the \ion{N}{1}\,$\lambda$1199.5 and \ion{Si}{3}\,$\lambda$1206 
profiles shown in the middle and lower panels, respectively,
of Figure~\ref{fig:mrk876_stack}.  The upper panel shows the same 
\ion{H}{1} two-component structure seen in previous sightlines; the
total \ion{H}{1} column density is N(\ion{H}{1})=(2.3$\pm$0.2)$\times$
10$^{19}$\,cm$^{-2}$.

\placefigure{fig:mrk876_stack}

The first thing to note from Figure~\ref{fig:mrk876_stack} is that 
both Galactic and Complex~C 
\ion{Si}{3}\,$\lambda$1206 absorption features
are clearly saturated and blended.  We can do
little more than set an unrestrictive lower limit on
[\ion{Si}{3}/\ion{H}{1}].  Integrating the line profile over the range
$-$250$\le v_{\rm LSR}\le$$-$95\,km\,s$^{-1}$ yields the equivalent
width (column 5) and apparent column densities (column 7) listed in
the final entry of Table~\ref{tbl:features}.  
These data imply
\begin{eqnarray}
{[\rm Si~III/H~I]} & > & -1.30\pm 0.04. \nonumber
\end{eqnarray}
\noindent
Unlike
sulfur, silicon is easily depleted onto dust, so an upward correction
of 0.3$-$1.0\,dex would not be unexpected (Savage \& Sembach 1996).  Both
ionization and line saturation effects act in the direction of increasing
[\ion{Si}{3}/\ion{H}{1}], although it should be noted that \ion{Si}{3} is not
expected to coincide spatially with the \ion{H}{1}.

The middle panel of Figure~\ref{fig:mrk876_stack} is perhaps
more intriguing.
While the data are noisy, there is clear evidence for an
extended ``blue'' wing to the 
saturated Galactic \ion{N}{1}\,$\lambda$1199.5
absorption feature, coinciding with the position of the Complex~C 
\ion{H}{1} emission (upper panel).  The inferred apparent column density
(Table~\ref{tbl:features}), coupled with the Effelsberg 
N(\ion{H}{1}), yields a measure of the nitrogen abundance for
Complex~C of
\begin{eqnarray}
{[\rm N~I/H~I]} & = & -1.41\pm 0.08. \nonumber
\end{eqnarray}
\noindent
In the absence of any 
ionization correction, this inferred metallicity is surprisingly low
in comparison with the previous
sightlines discussed here (\S\S~\ref{mrk290}-\ref{mrk501}).  A grid
of Cloudy models (see \S~\ref{mrk290})
was constructed to explore the magnitude of any potential
ionization correction.  For N(\ion{H}{1})$>$10$^{19}$\,cm$^{-2}$ and 
density n$_{\rm H}$$>$0.01\,cm$^{-3}$, we find
[\ion{N}{1}/\ion{H}{1}]= (0.6$-$1.0)\,[N/H].  For n$_{\rm
H}$$>$0.1\,cm$^{-3}$, [\ion{N}{1}/\ion{H}{1}] is equivalent to [N/H], to within
10\%.  Thus, for a wide range of ionization and density conditions, 
the ratio of
neutral nitrogen to hydrogen is a very good representation of the true
nitrogen metallicity.  This result is also insensitive to variations
in the shape of the spectrum of the incident radiation
field.
Since nitrogen is not depleted onto dust (Savage \& Sembach 1996) and the 
\ion{N}{1}\,$\lambda$1199.5 is not severely saturated, 
the current constraint upon
[\ion{N}{1}/\ion{H}{1}] appears robust.\footnote{This value is 
consistent with the
\ion{N}{1}\,$\lambda$1134.1 FUSE analysis of Murphy \etal\ (2000), who
found 
N(\ion{N}{1}\,$\lambda$1134.1)=(1.6$\pm$0.6)$\times$10$^{14}$\,cm$^{-2}$, 
in comparison with our value
N(\ion{N}{1}\,$\lambda$1199.5)=(1.0$\pm$0.1)$\times$10$^{14}$\,cm$^{-2}$.}

Since we expect that [\ion{N}{1}/\ion{H}{1}]
should trace [N/H], we are left with three possibilities for the 
low [\ion{N}{1}/\ion{H}{1}] metallicity inferred along this
sightline:  (i) either there are significant and unexpected
ionization corrections for higher-ion species of nitrogen, without 
a consequent correction for ionized hydrogen; (ii) the nucleosynthetic
history of Complex~C nitrogen is significantly different from that of
its sulfur enrichment; or (iii) our inferred \ion{N}{1} column density
is underestimated by approximately a factor of two.  
In the absence of proof to the contrary concerning (iii), we
believe that option (ii) is most likely, since nitrogen and sulfur have 
different nucleosynthetic origins.  Sulfur is produced in equal measure 
by Type Ia and Type II supernovae (Gibson, Loewenstein \& Mushotzky 1997),
while nitrogen is primarily produced by hot-bottom
burning in thermally-pulsating asymptotic giant branch (AGB) stars
(Gibson \& Mould 1997).  Primary
nitrogen might be a byproduct of primordial metallicity, high-mass
stars, but it is not the dominant nitrogen production mechanism.  If the
Mrk~876 sightline probes a region of Complex~C that, for whatever reason,
is free of AGB stellar ``pollution'', a significantly lower [N/S] or [N/Fe]
might be expected.  The Murphy \etal\ (2000) analysis of the 
FUSE Early Release Observations found [\ion{N}{1}/\ion{Fe}{2}]=$-$0.74,
in agreement with our (current) favored interpretation of this low
S/N STIS data.

\section{Discussion}
\label{summary}

We have undertaken a self-consistent GHRS and STIS 
analysis of five different probes of HVC
Complex~C.  It has been suggested that Complex~C represents the best example
of the predicted infalling low-metallicity gas responsible for fueling the bulk
of star formation in the Galaxy.  Such infalling gas is a necessary component
of Galactic chemical evolution models, in order to reconcile the paucity of
observed low-metallicity stars in the solar neighborhood with those predicted
upon theoretical grounds.

Based solely upon the Mrk~290 sightline, we confirm the earlier analysis of
Wakker \etal\ (1999a,b), who found [S/H]=$-$1.05$\pm$0.12 and support their
low-metallicity claim.  However, both the Mrk~817 and Mrk~279 sightlines
appear to probe significantly more metal-rich gas, at odds with the Mrk~290
result.  For Mrk~817, we found [\ion{S}{2}/\ion{H}{1}]=$-$0.48$\pm$0.06, a
factor of four higher metallicity than that found toward Mrk~290,
in keeping with metallicities observed in several other HVCs 
(Lu \etal\ 1998; Richter \etal\ 1999; Gibson \etal\ 2000).  For Mrk~279, 
[\ion{S}{2}/\ion{H}{1}]=$-$0.36$\pm$0.18, a factor of five higher than that
toward Mrk~290.  Despite the detection of 
\ion{Fe}{2}\,$\lambda\lambda$1122,1144 and \ion{Si}{2}\,$\lambda$1020 along
this line of sight, we cannot rule out the possibility that our 
\ion{S}{2}\,$\lambda$1250 feature is actually Ly$\alpha$ intrinsic to
Mrk~279.  The poorly constrained continuum near the expected
Mrk~501 S~II features allowed us to place only approximate upper limits to
[\ion{S}{2}/\ion{H}{1}].  Finally, for Mrk~876, we were able to
place limits on the nitrogen abundance along this Complex~C
sightline.  Our result, [\ion{N}{1}/\ion{H}{1}]=$-$1.41$\pm$0.08, suggests 
that scaled-solar abundance ratios are not encountered in Complex~C,
a suggestion supported by
Murphy et~al.'s (2000) conclusion that [\ion{N}{1}/\ion{Fe}{2}]$<$0.  Based on
[\ion{N}{1}/\ion{H}{1}] from Mrk~876 and [\ion{S}{2}/\ion{H}{1}] from Mrk~290
and 817, 
it would appear that [\ion{N}{1}/\ion{S}{2}]=$-$0.6$\pm$0.4,
suggestive of mild 
$\alpha$-element enrichment from supernovae, with only marginal
nitrogen pollution from hot-bottom burning in intermediate mass stars.

Table~\ref{tbl:summary} summarizes the extant GHRS and STIS data pertaining
to HVC Complex~C.  We reference all of our results to the \ion{H}{1} column
densities provided by the Effelsberg spectra of Wakker \etal\ (2001). 
While it is tempting to attach a $\pm$0.2\,dex systematic uncertainty to the
abundances due to the resolution mismatch, counter-examples do exist in the
literature which show that significantly larger variations in column might be
encountered at the arcsecond level.  We emphasize that our results are
limited by the 9$^\prime$ beam size \ion{H}{1} data used here. We will 
revisit our conclusions once higher spatial resolution 21\,cm data become 
available.

\placetable{tbl:summary}

We attempted to reconcile the higher inferred sulfur abundances along the
Mrk~817 and 279 sightlines with that seen toward Mrk~290, by invoking an
ionization model similar to that of Wakker \etal\ (1999a).  
The predicted H$\alpha$ emission measures are unrealistically high,
1200 and 1700\,mR, respectively.  We could also construct unrealistic
ionization models in which sulfur remains in the \ion{S}{2} state, even in
regions of the cloud in which all of the hydrogen is ionized. 
This would bring both the Mrk~817 and Mrk~279 sightlines into 
$\sim$1$\sigma$ agreement with the lower metallicity Mrk~290 data.  The
predicted H$\alpha$ emission measures, I(H$\alpha$)$\approx$900\,mR, 
are substantially lower, although still fairly high.
It is also quite possible that the other sightlines have different
density distributions, which introduces still more uncertainties
into an estimate of the expected I(H$\alpha$).
The remaining (large) 
uncertainties in the \ion{H}{1} column density determination
at arcsecond levels suggest that
detailed discussion of ionization conditions is not fruitful at this stage.

The Mrk~290 sightline through HVC Complex~C does support the suggestion that 
this cloud may be low-metallicity Galactic fuel.  However, our analysis of
the nearby Mrk~817 and and Mrk~279 sightlines does not, nor does the iron 
abundance found in the FUSE analysis of Mrk~876 (Murphy \etal\ 2000).  
On the other hand, the analysis of Richter \etal\ (2001) of the sightline 
toward PG~1259+593 indicates an [O~I/H~I] = $-$1.03$\pm$0.34, an 
abundance determination that is robust to ionization or depletion 
corrections. Our preliminary analysis of three O~I absorption profiles
toward Mrk~817 in the FUSE band yields [O~I/H~I] $= -0.88$ to $-0.65$, 
assuming the same Grevesse \& Noels (1993) solar oxygen abundence 
O/H $=7.4\times 10^{-4}$ as Richter et al.\ (2001); adopting
the revised Holweger (2001) solar oxygen abundance results in a 
Complex~C oxygen abundance of [O~I/H~I] $= -0.73$ to $-0.51$.  In 
other words, \it the oxygen abundance along the Mrk~817 sightline is
$\sim$0.2$-$0.3$\times$ solar. \rm
If Complex~C has a 
single characteristic metallicity, a compelling case does not exist to 
reconcile all observations with either the low-metallicity Galactic fuel 
scenario ([S/H]$\approx$$-$1.0), or the higher-metallicity Galactic 
waste/disrupted processed gas scenario (with [S/H]$\simgt$$-$0.5).  

In the absence of a compelling argument for retaining the above
assumption of mono-metallicity, it is worth considering the hypothesis that the
observed differences may simply reflect internal abundance variations within 
the HVC.  A useful example of such variations is provided by the SMC.  First,
the mean metallicity of the SMC is $\sim$0.2 times solar (Gibson \etal\ 2000;
Table~6), which is comparable to that of Complex~C.  Second, its linear extent
is also comparable to that of Complex~C, assuming the latter is at a distance
of $\sim$10\,kpc (Wakker \etal\ 1999a,b).\footnote{The \ion{H}{1} mass of the
SMC is of course much greater than that of Complex~C, by approximately two
orders of magnitude (Wakker \& van Woerden 1991).}  Our analysis of the
Mrk~290, 817, and 279 sightlines, in conjunction with that of Mrk~876 by Murphy
\etal\ (2000), implies the presence of Complex~C abundance differences on the 
order of factors of 3$-$4 over scales of 1$-$2\,kpc, the typical impact 
parameter between these four probes.  For comparison, 
both F-type supergiants (Russell \& Bessell 1989) and B-stars (Rolleston
\etal\ 1999) in the SMC show abundance variations on the order of a factor of
$\sim$3.  In other words, a nearby system of both comparable size and
metallicity to Complex~C shows abundance variations at the level of
$\sim$0.5\,dex, similar to the magnitude of abundance variations implied by
Table~\ref{tbl:summary} and Murphy \etal\ (2000).  This argument does not prove
that abundance variations in Complex~C exist, but is simply provided to show
that they cannot be excluded at this point, and indeed may not be 
unexpected.

Between latitudes $+$15$\simlt$$b$$\simlt$$+$30, it has long
been recognised (e.g. Davies 1972; Verschuur 1975, and
references therein) that Complex~C connects
smoothly (in terms of kinematics) to what is now dubbed 
the Outer Arm Complex (adopting the Wakker \& van~Woerden 1991 nomenclature)
of the Milky Way (in the Galaxy's first quadrant).
The models of Davies and Verschuur place Complex~C (and indeed, many of
the northern HVCs) in the outer Galaxy, either as a high-scale height
extension of the Outer Arm (where the maximum in the disk's warp
occurs) or a separate spiral arm.  In contrast, Haud (1988) suggests
the Milky Way is a polar ring galaxy, with the Magellanic Stream and
Complex~C part of a lengthy, near-continuous feature, surrounding
the Galaxy.  Metallicity differences between the two HVCs
(compare Gibson et~al. 2000 with the results presented here) likely make the
Haud scenario untenable (as does the fact that the Stream and Complex~C
have significantly different position angles on the sky); 
on the other hand, the metallicity of Complex~C
does match that of the outer disk (e.g. Gibson et~al. 2002; Figure~2),
as one might expect if the connection to the Outer Arm was true.
It should be stressed though that this metallicity match may simply be
coincidental.  Regardless, 
a revisit to the classic Davies and Verschuur
models appears timely.  We are currently
pursuing 3-dimensional hydrodynamical
simulations of the Outer Arm Complex and Complex~C, in an attempt to unravel
the kinematics in this region of the Galaxy.

At this point, we urge that caution be exercised in identifying the HVC
Complex~C as the long-sought low-metallicity infalling gas.  Future 
observations with FUSE and HST/STIS (e.g. Mrk~205 by Bowen et~al.)
of Complex~C should clarify many of the unanswered questions raised here.

\acknowledgments

We wish to express our gratitude to Bart Wakker, who provided feedback above
and beyond the call of duty.  Special thanks are also due Peter Kalberla who 
provided \ion{H}{1} spectra in advance of their publication.  JMS, MLG, and
BKG acknowledge the financial support at the University of Colorado
through the NASA Long-Term Space Astrophysics Program (NAG5-7262) and the 
FUSE Science Team (NAS5-32985).
We also thank the HST program GO-0653.01-95A for use of the GHRS spectra.

\clearpage

\figcaption[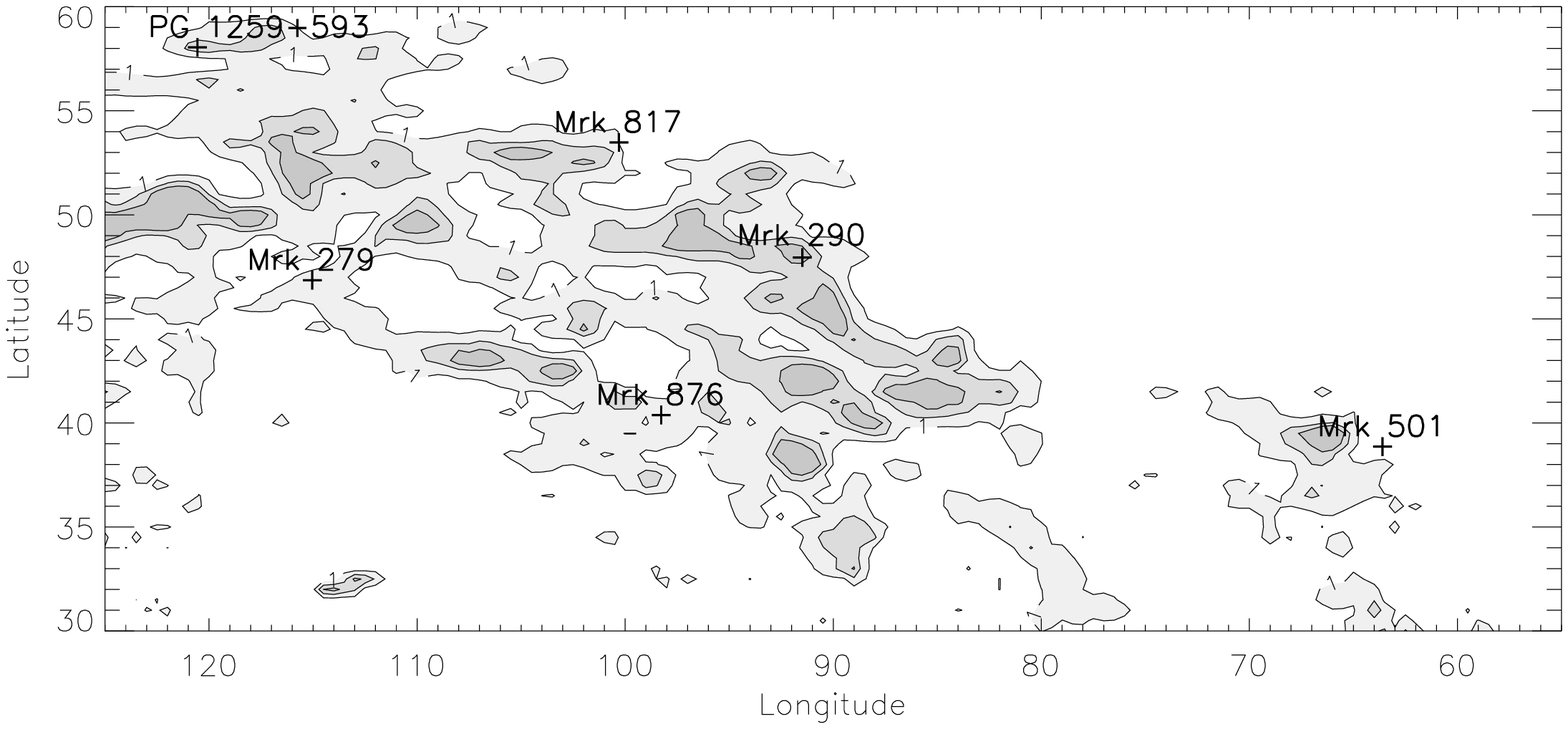]{
\baselineskip=14pt
Contours of HVC neutral hydrogen column density
N(\ion{H}{1}) centered on Complex~C over the velocity range $-250<v_{\rm
LSR}<-110$\,km\,s$^{-1}$.
Contour levels are $2\times 10^{19}$\,cm$^{-2}$, with the outermost level
N$_{\rm H\,I}=1\times 10^{19}$\,cm$^{-2}$.  The \ion{H}{1}
data were taken from the
Leiden-Dwingeloo Survey (Hartmann \& Burton 1997).  The five background probes
discussed here (Mrk~290, 817, 279, 501, and 876) are labeled.  A more detailed
map of the area is shown in Wakker (2001).
\label{fig:map}}

\figcaption[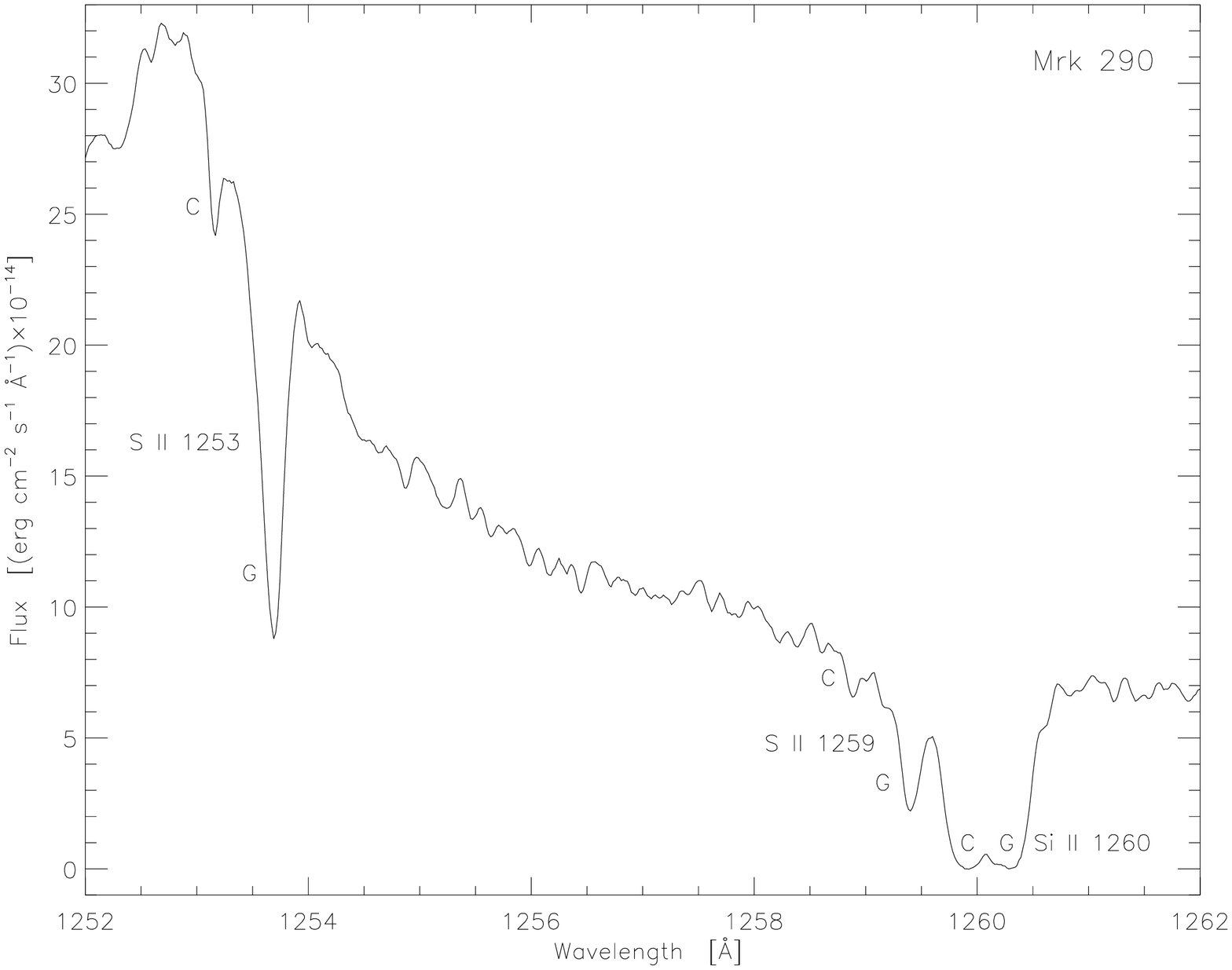]{
\baselineskip=14pt
HST GHRS spectrum of Mrk~290 taken with the G160M
grating.  The co-added
spectrum has been smoothed by the post-COSTAR, GHRS, large science aperture,
line spread function (Gilliland 1994).  The Galaxy (G) and Complex~C (C) are
seen in \ion{S}{2}\,$\lambda$1253, \ion{S}{2}\,$\lambda$1259, and 
\ion{Si}{2}\,$\lambda$1260, and labeled accordingly.  The expected
\ion{S}{2}\,$\lambda$1250 feature is obscured by intrinsic Ly$\alpha$
absorption.
\label{fig:mrk290_ghrs}}

\figcaption[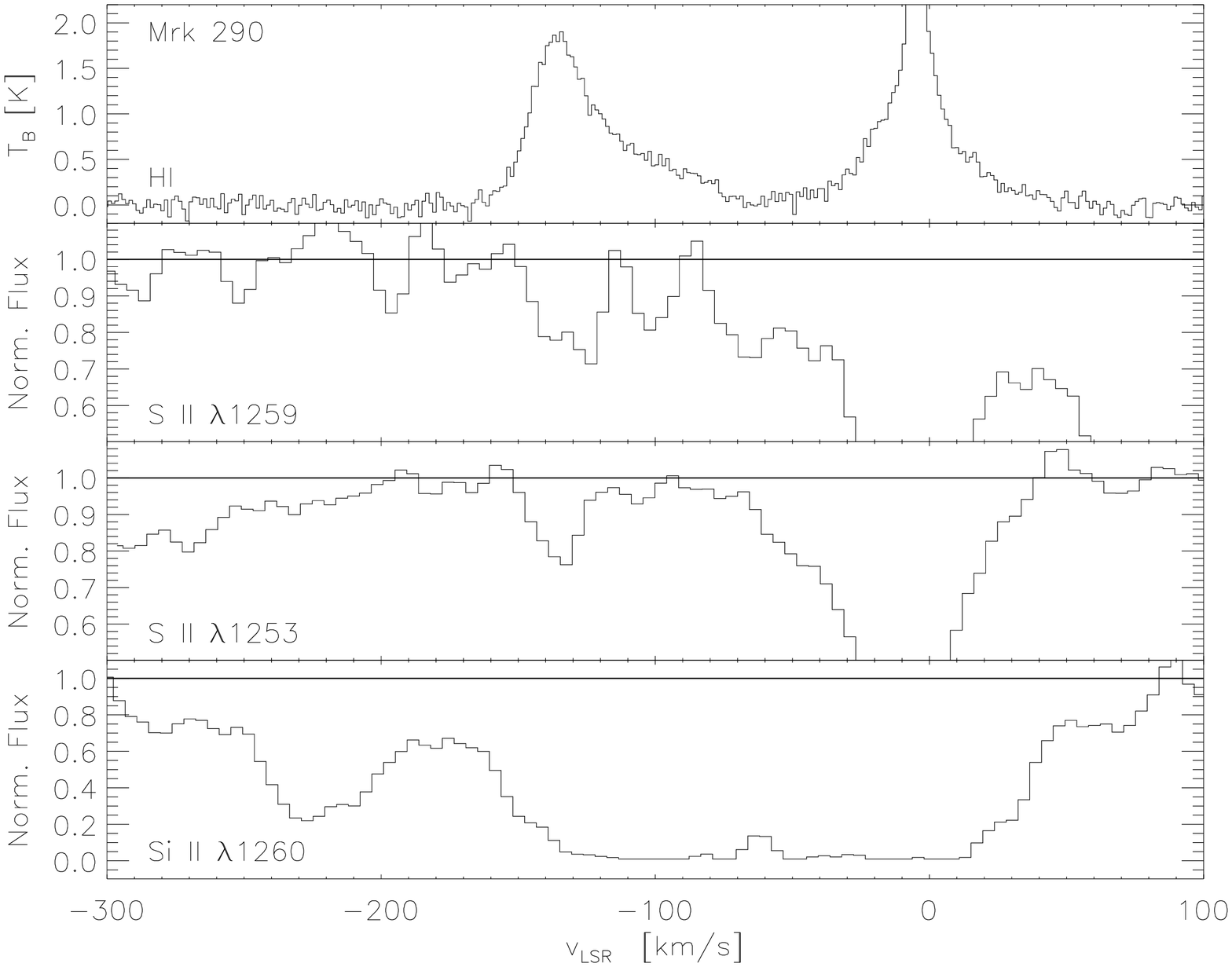]{
\baselineskip=14pt
Velocity stack showing \ion{H}{1} (upper panel),
\ion{S}{2} (middle panels), and \ion{Si}{2} (lower panel)
absorption features along the
Mrk~290 sightline.  The Effelsberg \ion{H}{1} data are from Wakker \etal\ 
(2001).  The raw (\ie, not smoothed by the post-COSTAR line spread 
function)
GHRS data are shown in the bottom three panels.  Centroid, equivalent widths,
and column densities for the Complex~C absorption features are listed in
Table~\ref{tbl:features}.
\label{fig:mrk290_stack}}

\figcaption[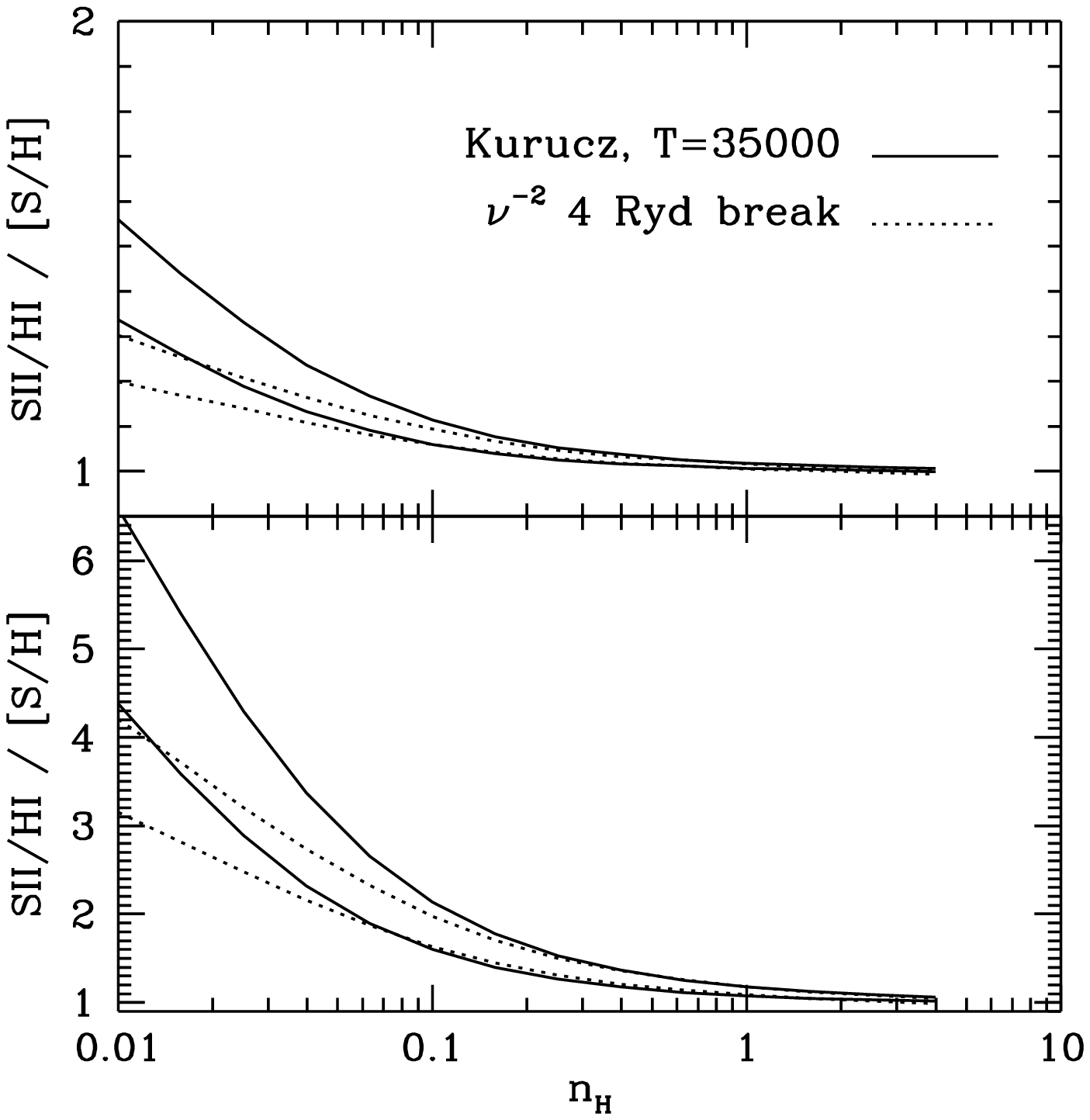]{
\baselineskip=14pt
Ionization correction to \ion{S}{2}/\ion{H}{1}
versus density $n_{\rm H}$.  All curves are based upon photoionization models
which assume a one-sided, normally-incident photon flux 
$\phi=10^{5.5}$\,photons\,cm$^{-2}$\,s$^{-1}$.  
Upper panel assumes plane-parallel slab with N(\ion{H}{1})=
10$^{20}$\,cm$^{-2}$.  Solid and dotted curves assume
spectral shape for ionizing radiation as labeled.  Upper curves
assume Z/Z$_{\odot}=0.1$.  Lower curves assume solar metallicity.
Lower panel is same as upper, except N(\ion{H}{1})=
10$^{19}$\,cm$^{-2}$.  
\label{fig:SII}}

\figcaption[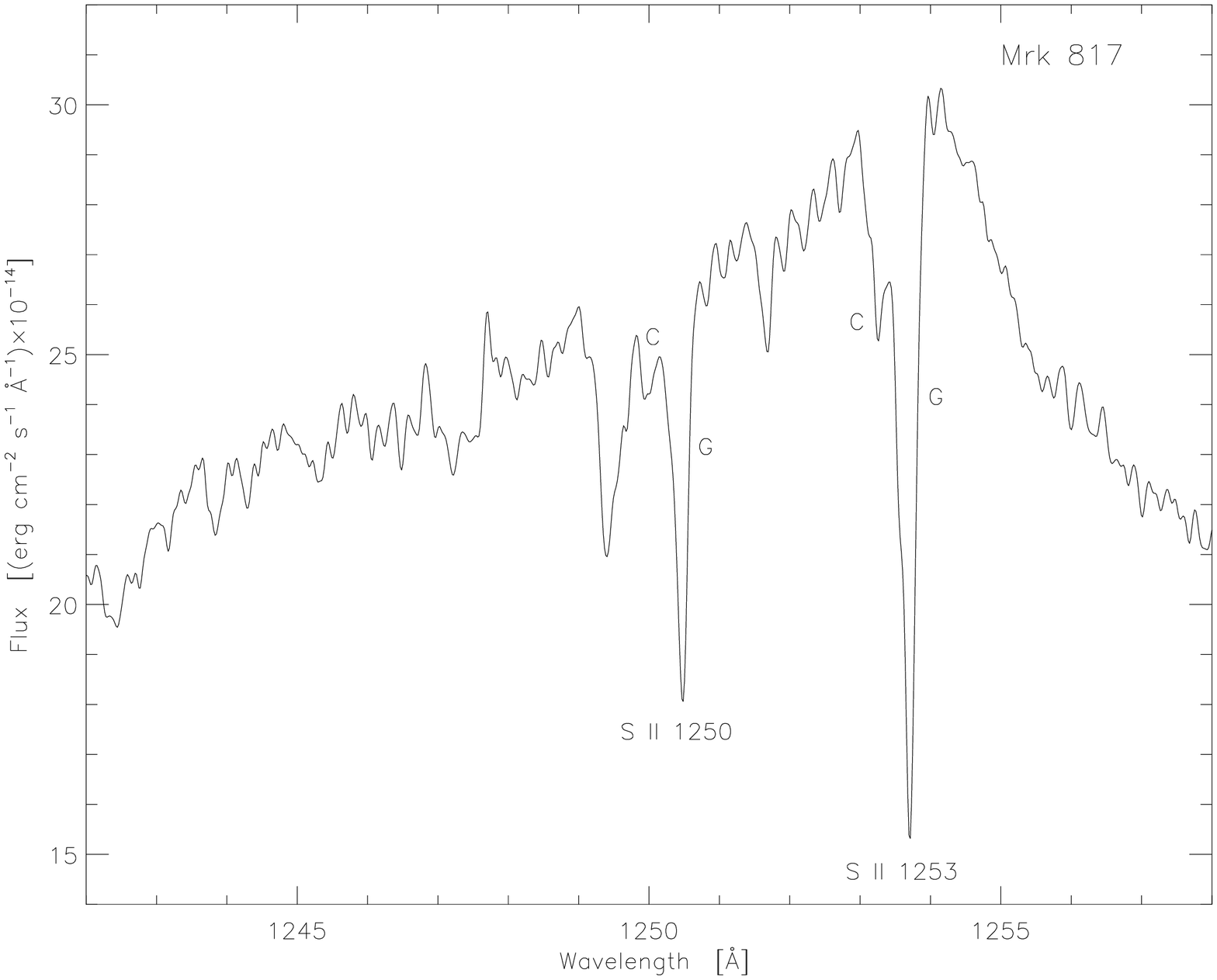]{
\baselineskip=14pt
HST GHRS spectrum of Mrk~817 taken with the G160M
grating.  The co-added
spectrum has been smoothed by the post-COSTAR, GHRS, large science aperture,
line spread function (Gilliland 1994).  The Galaxy (G) and Complex~C (C) are
seen in \ion{S}{2}\,$\lambda$1250 and \ion{S}{2}\,$\lambda$1253 and 
labeled accordingly.
\label{fig:mrk817_ghrs}}

\figcaption[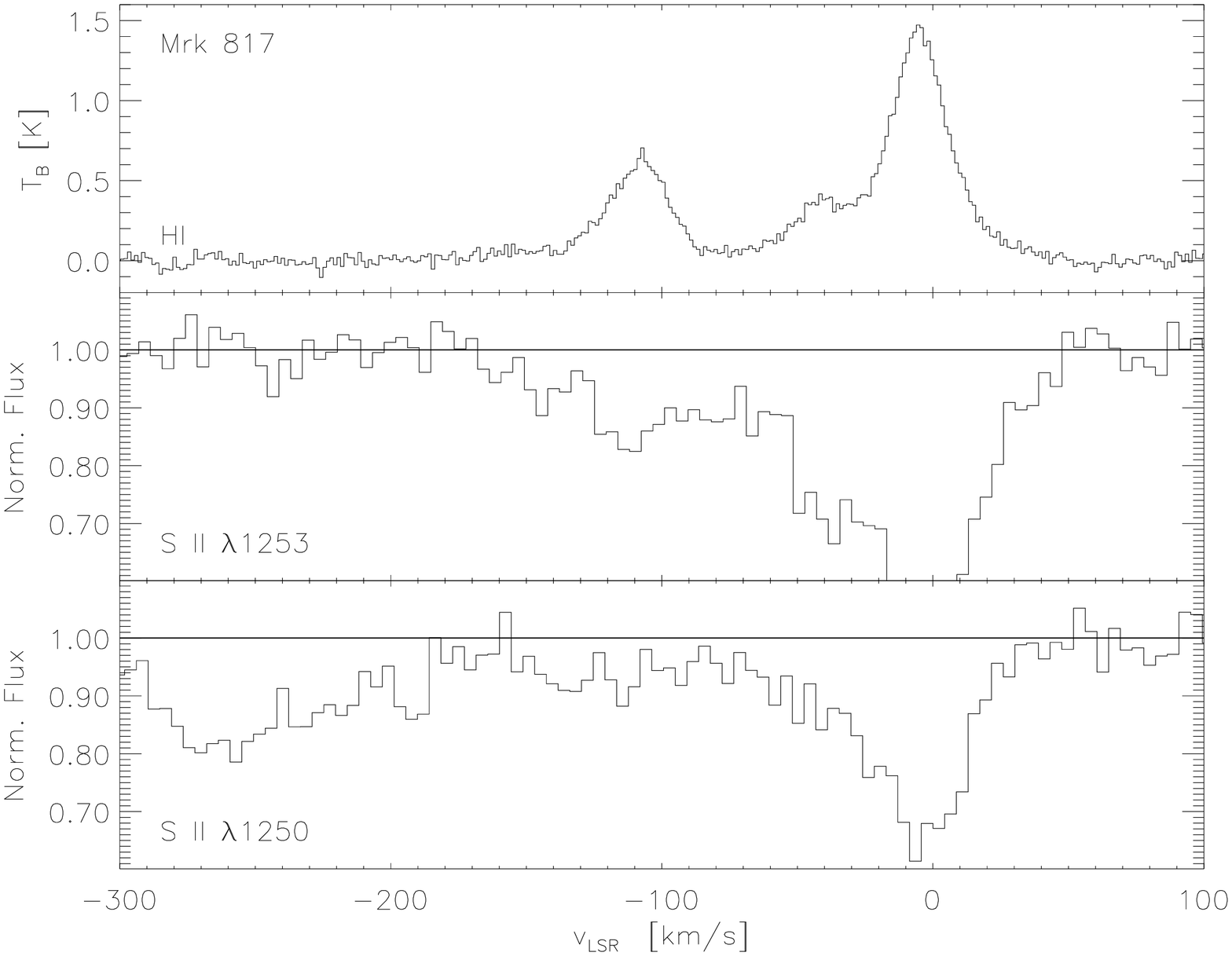]{
\baselineskip=14pt
Velocity stack showing \ion{H}{1} (upper panel),
\ion{S}{2}\,$\lambda$1253 (middle panel), and \ion{S}{2}\,$\lambda$1250
(lower panel) absorption features along the
Mrk~817 sightline.  The Effelsberg \ion{H}{1} data are from Wakker \etal\ 
(2001).  The raw (\ie, not smoothed by the post-COSTAR line spread 
function)
GHRS data are shown in the bottom two panels.  Centroid, equivalent widths,
and column densities for the Complex~C absorption features are listed in
Table~\ref{tbl:features}.
\label{fig:mrk817_stack}}

\figcaption[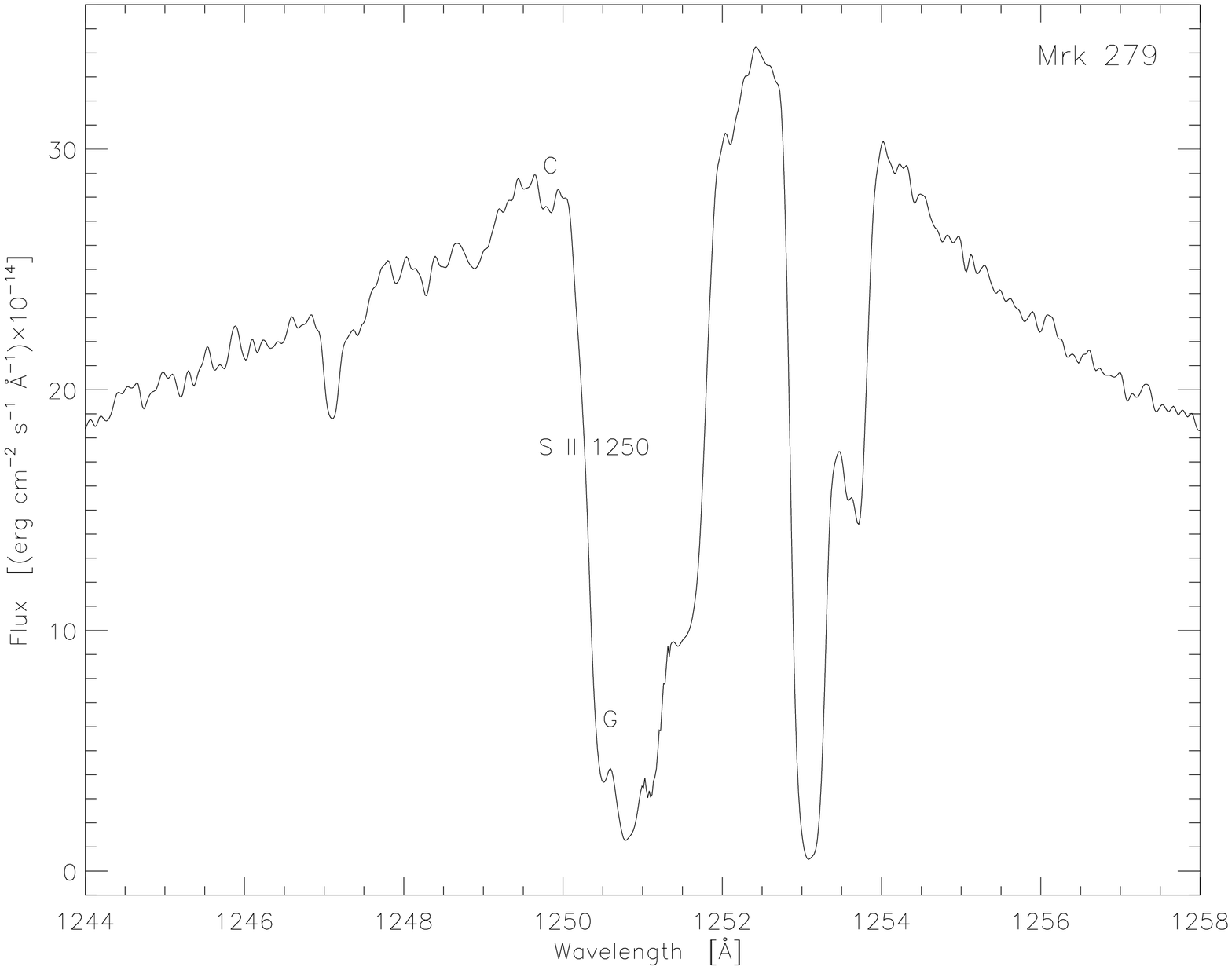]{
\baselineskip=14pt
HST GHRS spectrum of Mrk~279 taken with the G160M
grating.  The co-added
spectrum has been smoothed by the post-COSTAR, GHRS, large science aperture,
line spread function (Gilliland 1994).  The Galaxy (G) and Complex~C (C) are
seen in \ion{S}{2}\,$\lambda$1250 and labeled accordingly.
\label{fig:mrk279_ghrs}}

\figcaption[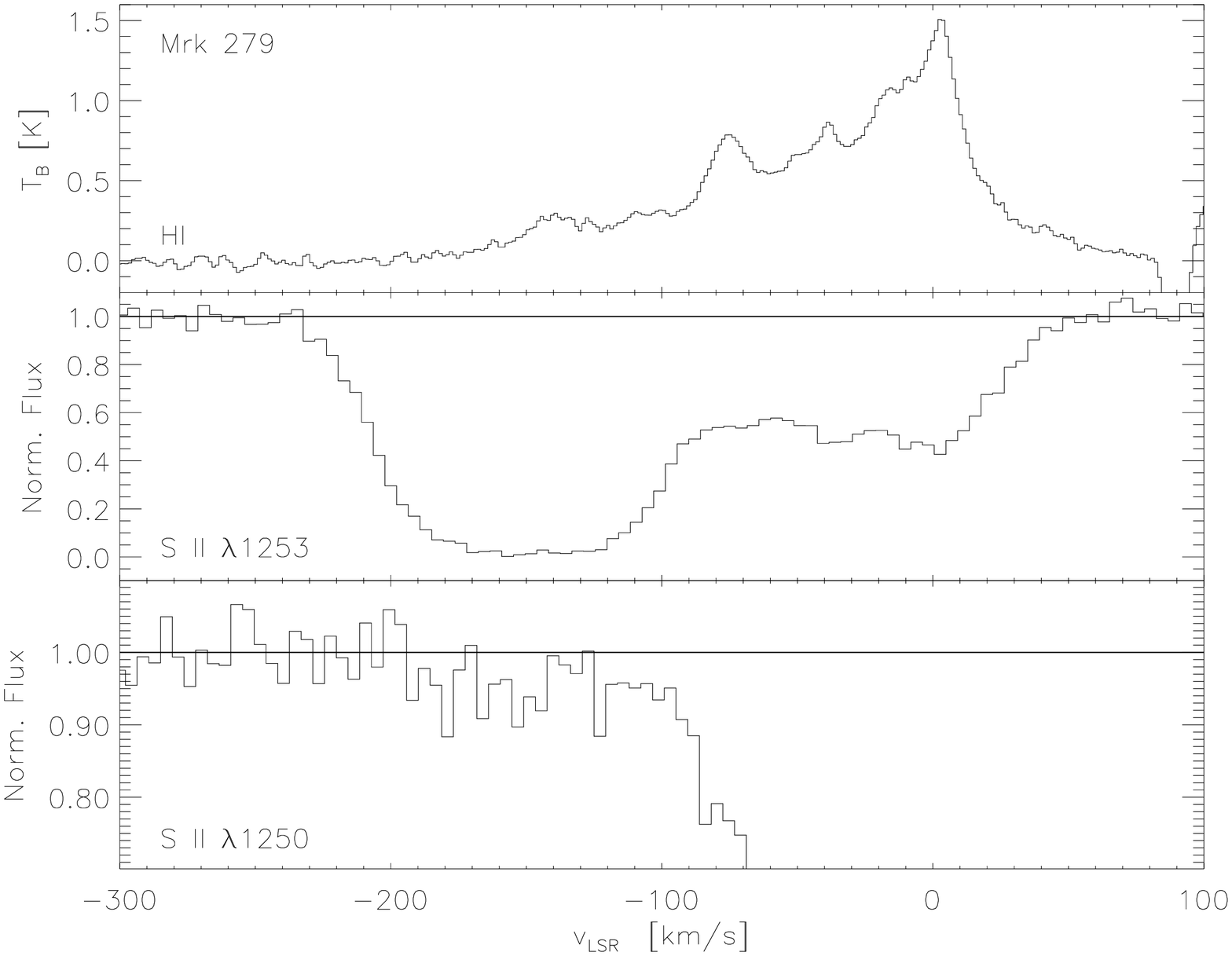]{
\baselineskip=14pt
Velocity stack showing \ion{H}{1} (upper panel),
\ion{S}{2}\,$\lambda$1253 middle panel), and \ion{S}{2}\,$\lambda$1250
(lower panel) absorption features along the
Mrk~279 sightline.  The Effelsberg \ion{H}{1} data are from Wakker \etal\ 
(2001).  The raw (\ie, not smoothed by the post-COSTAR line spread 
function)
GHRS data are shown in the bottom two panels.  Intrinsic Mrk~279
Ly$\alpha$ absorption obscures the expected
Complex~C \ion{S}{2}\,$\lambda$1253 absorption feature in the middle panel.
Centroid, equivalent widths,
and column densities for the Complex~C \ion{S}{2}\,$\lambda$1250
absorption feature are listed in
Table~\ref{tbl:features}.
\label{fig:mrk279_stack}}

\figcaption[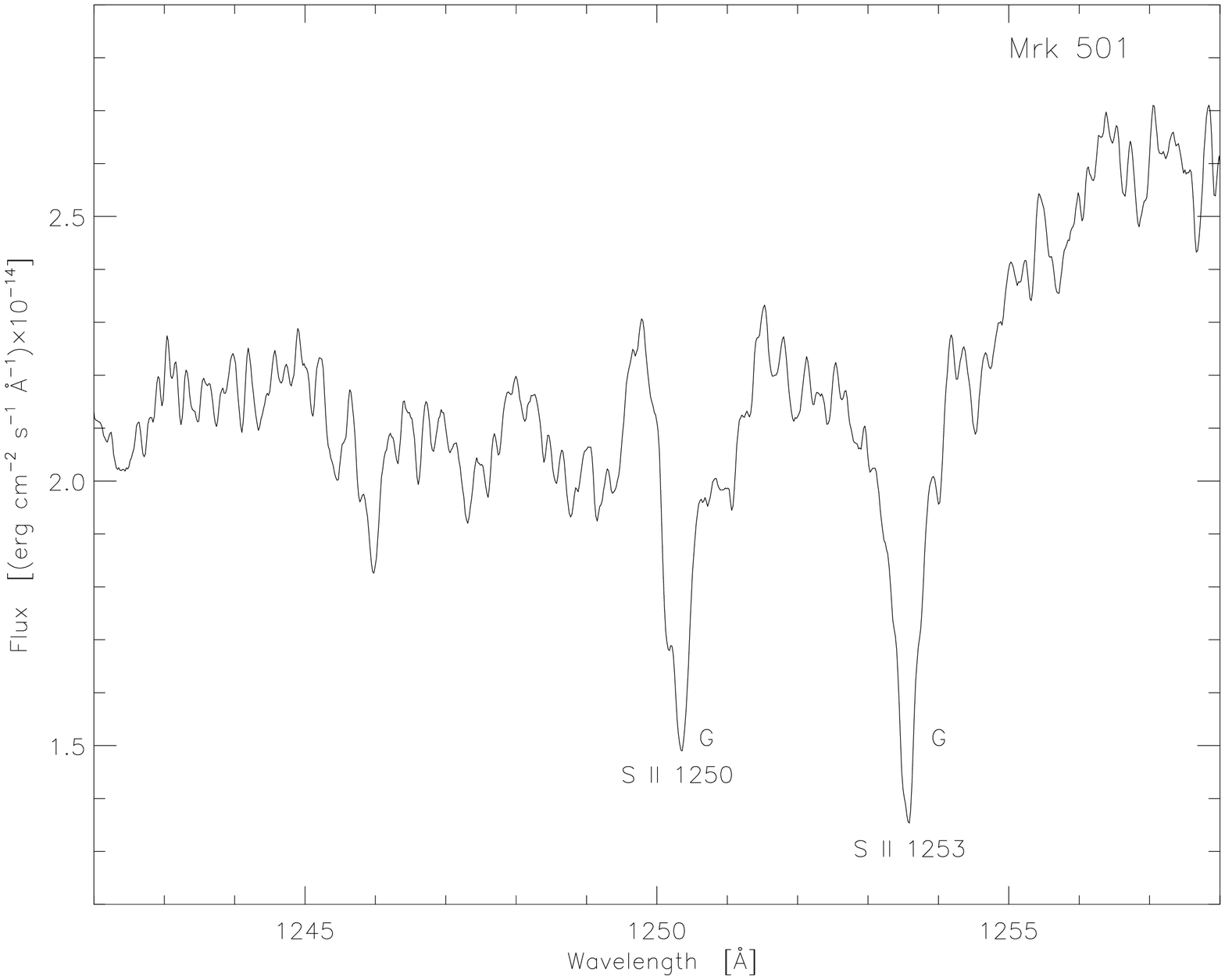]{
\baselineskip=14pt
HST GHRS spectrum of Mrk~279 taken with the G160M
grating.  The 
spectrum has been smoothed by the pre-COSTAR, GHRS, large science aperture,
line spread function (Gilliland 1994).  The Galaxy (G) is
seen in \ion{S}{2}\,$\lambda$1250 and \ion{S}{2}\,$\lambda$1253, 
and labeled accordingly.  The continuum in the vicinity of both Galactic
features is poorly determined, complicating the identification of any potential
Complex~C absorption features.
\label{fig:mrk501_ghrs}}

\figcaption[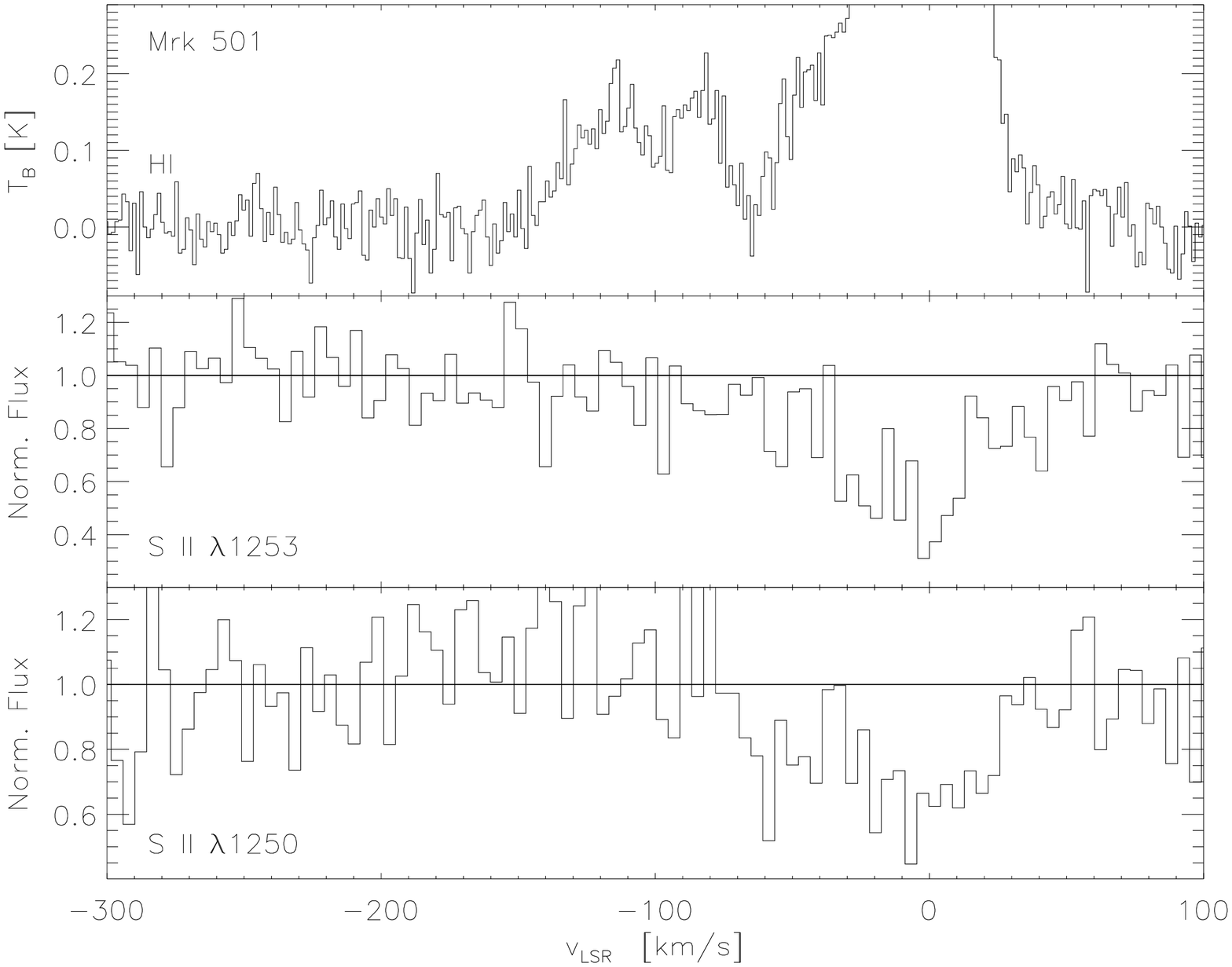]{
\baselineskip=14pt
Velocity stack showing \ion{H}{1} (upper panel),
\ion{S}{2}\,$\lambda$1253 (middle panel), and \ion{S}{2}\,$\lambda$1250
(lower panel) absorption features along the
Mrk~501 sightline.  The Effelsberg \ion{H}{1} data are from Wakker \etal\ 
(2001).  The raw (\ie, not smoothed by the pre-COSTAR line spread 
function)
GHRS data are shown in the bottom two panels.  Absorption due to 
Complex~C is not seen unambiguously in either \ion{S}{2} line, yielding only
a loose constraint on the upper limit on the metallicity along this sightline
of [\ion{S}{2}/\ion{H}{1}]$\simlt$$-$0.1.
\label{fig:mrk501_stack}}

\figcaption[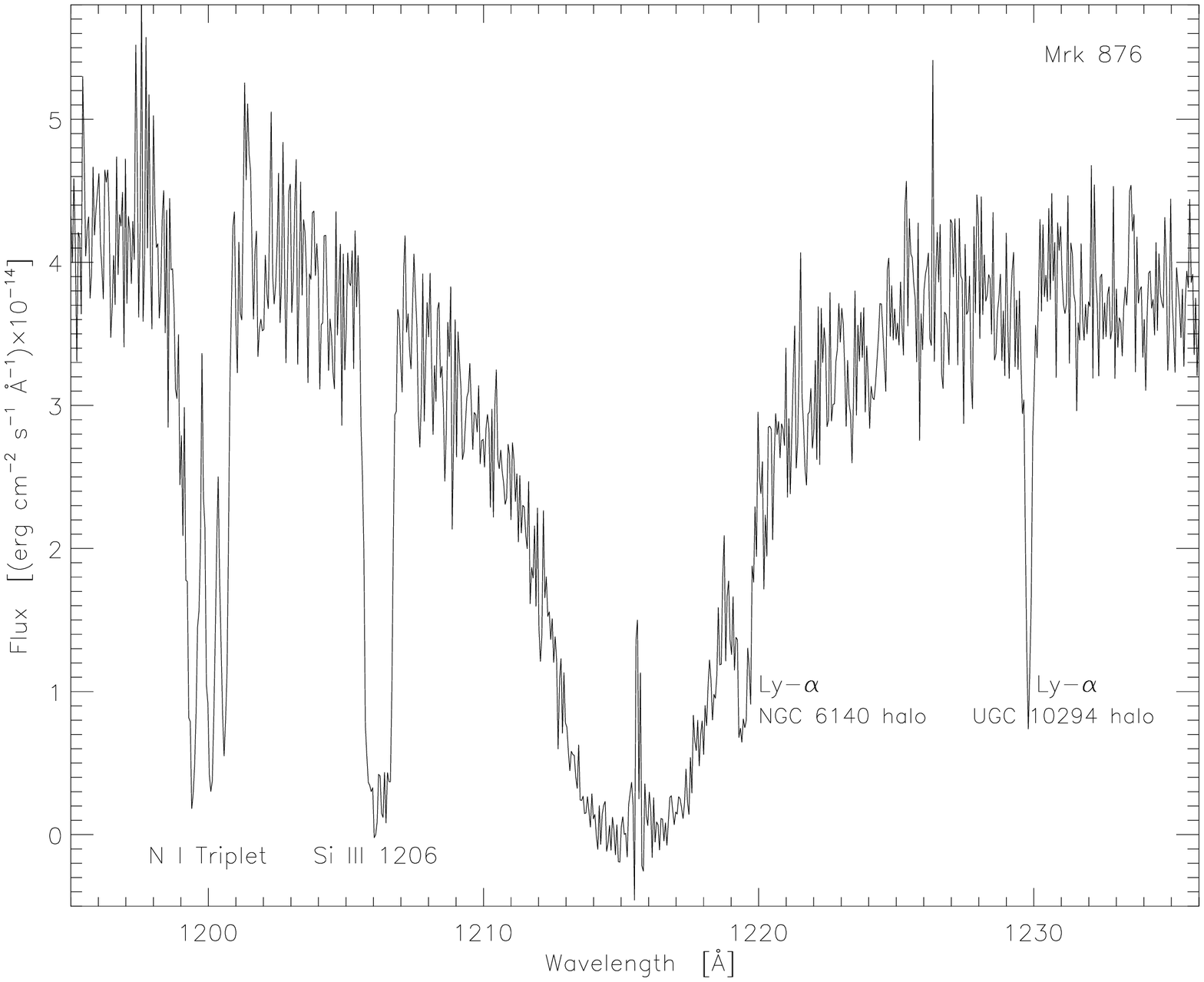]{
\baselineskip=14pt
HST STIS spectrum of Mrk~876 taken with the G140M
grating.  Complex~C
absorption is seen in the blueward wings of the saturated Galactic \ion{N}{1}
and \ion{Si}{3} lines.  Galactic Ly$\alpha$ dominates the spectrum.  Two
intervening galactic halos are detected in Ly$\alpha$: (i) the extended
halo of NGC~6140, at $v\approx +930$\,km\,s$^{-1}$ (impact parameter of
$\sim$180\,$h_{70}^{-1}$\,kpc); 
and (ii) the extended halo of UGC~10294, at $v\approx
3490$\,km\,s$^{-1}$ (impact parameter of $\sim$270\,$h_{70}^{-1}$\,kpc) 
and will not be discussed further.
\label{fig:mrk876_stis}}

\figcaption[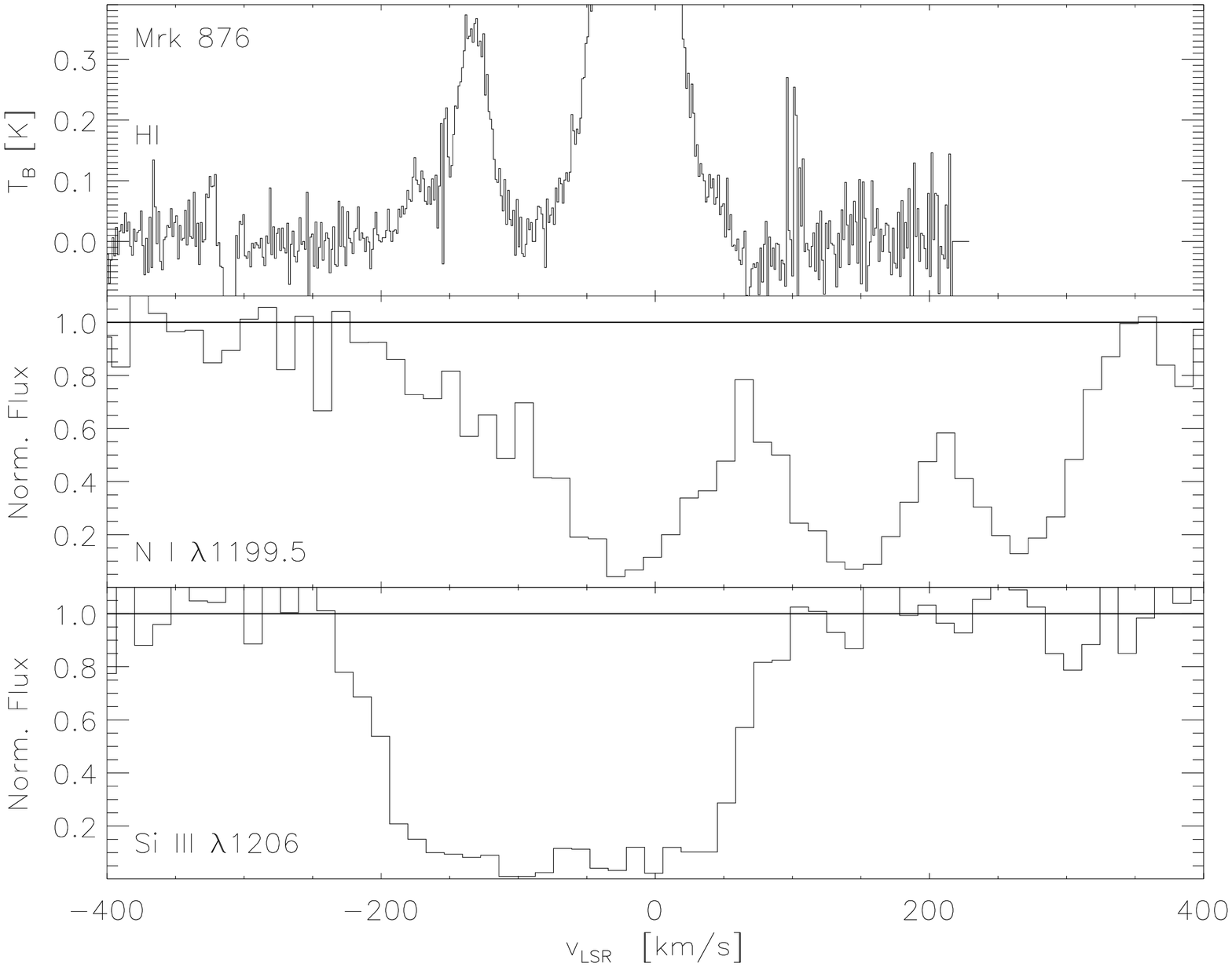]{
\baselineskip=14pt
Velocity stack showing \ion{H}{1} (upper panel),
\ion{N}{1}\,$\lambda$1199.5 (middle panel), and \ion{Si}{3}\,$\lambda$1206
(lower panel) absorption features along the
Mrk~876 sightline.  The Effelsberg \ion{H}{1} data are from Wakker \etal\ 
(2001).  N(\ion{N}{1}\,$\lambda$1199.5) inferred from the blueward
absorption ``wing'' of the Galactic \ion{N}{1}$\lambda$1199.5 feature
is consistent with 
N(\ion{N}{1}\,$\lambda$1134.1), derived by Murphy \etal\ (2000).
Centroids, equivalent widths,
and column densities for the Complex~C \ion{N}{1}\,$\lambda$1199.5
and \ion{Si}{3}\,1206 absorption features are listed in 
Table~\ref{tbl:features}.
\label{fig:mrk876_stack}}

\clearpage

\begin{deluxetable}{llclrccrr}
\tabletypesize{\footnotesize}
\tablecaption{HST Observations
\label{tbl:obs}}
\tablewidth{0pt}
\tablehead{
\colhead{PID/PI\tablenotemark{a}} &
\colhead{Dataset\tablenotemark{b}} &
\colhead{Date} &
\colhead{Target} &
\colhead{t$_{\rm exp}$ [s]} &
\colhead{Instrument} &
\colhead{Grating} &
\colhead{$\lambda_i$ [\AA]} &
\colhead{$\lambda_f$ [\AA]}
}
\startdata 
6590/Wakker & z3kh0104t & 01/10/97 & Mrk~290   &  1622.0 & GHRS & G160M & 1231.8 & 1269.0 \\
6590/Wakker & z3kh0105t & 01/10/97 & Mrk~290   &  2433.0 & GHRS & G160M & 1231.8 & 1269.0 \\
6590/Wakker & z3kh0107t & 01/10/97 & Mrk~290   &  2230.3 & GHRS & G160M & 1231.8 & 1269.0 \\
6593/Stocke & z3e70104t & 01/12/97 & Mrk~817   &  8870.4 & GHRS & G160M & 1222.6 & 1258.7 \\
6593/Stocke & z3e70106t & 01/12/97 & Mrk~817   &  8236.8 & GHRS & G160M & 1222.6 & 1258.7 \\
6593/Stocke & z3e70108t & 01/12/97 & Mrk~817   &  7907.3 & GHRS & G160M & 1222.6 & 1258.8 \\
6593/Stocke & z3e70304t & 01/17/97 & Mrk~279   &  5702.4 & GHRS & G160M & 1222.5 & 1258.7 \\
6593/Stocke & z3e70306t & 01/17/97 & Mrk~279   &  5702.4 & GHRS & G160M & 1222.6 & 1258.7 \\
6593/Stocke & z3e70308t & 01/17/97 & Mrk~279   &  5702.4 & GHRS & G160M & 1222.6 & 1258.7 \\
6593/Stocke & z3e7030at & 01/17/97 & Mrk~279   &  1419.3 & GHRS & G160M & 1222.6 & 1258.8 \\
3584/Stocke & z1a65204m & 02/28/93 & Mrk~501   & 29196.3 & GHRS & G160M & 1221.4 & 1257.6 \\
7295/Cot\'e & o4n30801o & 09/19/98 & Mrk~876   &  2298.0 & STIS & G140M & 1194.8 & 1249.2 \\
\enddata
\tablenotetext{a}{HST Proposal ID and Principal Investigator.}
\tablenotetext{b}{HST Archive dataset filename.}
\end{deluxetable}

\clearpage

\begin{deluxetable}{llll}
\tabletypesize{\footnotesize}
\tablecaption{Atomic Data\tablenotemark{a}
\label{tbl:atomic}}
\tablewidth{0pt}
\tablehead{
\colhead{ID} &
\colhead{$\lambda_\circ$\, [\AA]} &
\colhead{$f$} &
\colhead{A$_{{\rm X}_{\odot}}$\tablenotemark{b}}
}
\startdata 
\ion{N}{1}\,$\lambda$1199 & 1199.5496 & 0.1328  & $(1.122\pm 0.104)\times 10^{-4}$ \\
\ion{S}{2}\,$\lambda$1250 & 1250.578  & 0.00545 & $(1.862\pm 0.215)\times 10^{-5}$ \\
\ion{S}{2}\,$\lambda$1253 & 1253.805  & 0.0109  & $\qquad\qquad\;\;^{\prime\prime}$       \\
\ion{S}{2}\,$\lambda$1259 & 1259.518  & 0.0162  & $\qquad\qquad\;\;^{\prime\prime}$       \\
\ion{Si}{3}\,$\lambda$1206& 1206.500  & 1.669   & $(3.548\pm 0.164)\times 10^{-5}$ \\
\ion{Si}{2}\,$\lambda$1260& 1260.4221 & 1.007   & $\qquad\qquad\;\;^{\prime\prime}$ \\
\enddata
\tablenotetext{a}{Rest wavelengths and oscillator strengths from Morton (1991).}
\tablenotetext{b}{Solar (meteoritic) abundance from Table 2 of
Anders \& Grevesse (1989), for the element X (where X corresponds to N,
S and Si, here).}
\end{deluxetable}

\clearpage

\begin{deluxetable}{llccrcr}
\tabletypesize{\footnotesize}
\tablecaption{Complex~C Absorption Features
\label{tbl:features}}
\tablewidth{0pt}
\tablehead{
\colhead{Probe} &
\colhead{Line ID} &
\colhead{$\lambda_c$\tablenotemark{a}} &
\colhead{$\Delta v_{{\rm LSR}}$\tablenotemark{b}} &
\colhead{W$_\lambda$} &
\colhead{N$_{\tau=0}$\tablenotemark{c}} &
\colhead{N$_{\tau_v}$\tablenotemark{d}} \\
\colhead{} &
\colhead{} &
\colhead{[\AA]} &
\colhead{[km\,s$^{-1}$]} &
\colhead{[m\AA]} &
\colhead{[cm$^{-2}$]} &
\colhead{[cm$^{-2}$]}
}
\startdata
   Mrk~290   &   S~II\,$\lambda$1253   &   1253.2                             &
				           $-$170$\leftrightarrow$$-$85       &
				           23$\pm$3$\,\,\,$                   &
				           (1.61$\pm$0.21)$\times$10$^{14}$   &
                                           (1.65$\pm$0.17)$\times$10$^{14}$ \\
   Mrk~290   &   S~II\,$\lambda$1259   &   1258.9                             &
				           $-$170$\leftrightarrow$$-$85       &
				           37$\pm$5$\,\,\,$                   &
				           (1.70$\pm$0.23)$\times$10$^{14}$   &
                                           (1.81$\pm$0.24)$\times$10$^{14}$ \\
   Mrk~290   &  Si~II\,$\lambda$1260\tablenotemark{e} &   1259.9              &
				           $-$180$\leftrightarrow$$-$60       &
				          399$\pm$5$\,\,\,$                   &
				           (2.82$\pm$0.04)$\times$10$^{13}$   &
                                           $>$1.0$\times$10$^{14}$          \\
   Mrk~817   &   S~II\,$\lambda$1250   &   1249.9                             &
				           $-$140$\leftrightarrow$$-$80       &
				           13$\pm$3$\,\,\,$                   &
				           (1.81$\pm$0.42)$\times$10$^{14}$   &
                                           (1.83$\pm$0.29)$\times$10$^{14}$ \\
   Mrk~817   &   S~II\,$\lambda$1253   &   1253.3                             &
				           $-$140$\leftrightarrow$$-$80       &
				           27$\pm$3$\,\,\,$                   &
				           (1.89$\pm$0.21)$\times$10$^{14}$   &
                                           (1.90$\pm$0.14)$\times$10$^{14}$ \\
   Mrk~279   &   S~II\,$\lambda$1250   &   1249.8                             &
				           $-$180$\leftrightarrow$$-$90       &
				           18$\pm$3$\,\,\,$                   &
				           (2.50$\pm$0.70)$\times$10$^{14}$   &
                                           (2.48$\pm$0.68)$\times$10$^{14}$ \\
   Mrk~876   &    N~I\,$\lambda$1199.5 &   1198.9                             &
				           $-$210$\leftrightarrow$$-$95       &
				          134$\pm$15                          &
				           (0.65$\pm$0.07)$\times$10$^{14}$   &
				           (1.01$\pm$0.13)$\times$10$^{14}$ \\
   Mrk~876   & Si~III\,$\lambda$1206   &   1205.8                             &
				           $-$250$\leftrightarrow$$-$95       &
				          363$\pm$12                          &
				           (1.68$\pm$0.06)$\times$10$^{13}$   &
                                           $>$4.1$\times$10$^{13}$          \\
\enddata
\tablenotetext{a}{Line centroid.}
\tablenotetext{b}{Velocity range over which the spectral
line integration was applied.}
\tablenotetext{c}{Inferred column density, under the assumption that
the line in question is optically thin.}
\tablenotetext{d}{Inferred column density, employing the $\tau_v$ technique
of Sembach \& Savage (1992), neglecting continuum placement
uncertainties.}
\tablenotetext{e}{Blended with Complex~C
\ion{Fe}{2}\,$\lambda$1260 absorption.}
\end{deluxetable}

\clearpage

\setlength{\topmargin}{13.0mm}
\thispagestyle{empty}
\begin{deluxetable}{lcccccccccc}
\rotate
\tabletypesize{\footnotesize}
\tablecaption{Summary of Complex~C Abundance Determinations\tablenotemark{a}
\label{tbl:summary}}
\tablewidth{0pt}
\tablehead{
\colhead{Probe} &
\colhead{N(S~II)} &
\colhead{N(N~I)} &
\colhead{N(Si~II)} &
\colhead{N(Si~III)} &
\colhead{N(H~I)\tablenotemark{b}} &
\colhead{[S~II/H~I]} &  
\colhead{[N~I/H~I]} &
\colhead{[Si~II/H~I]} &
\colhead{[Si~III/H~I]} &
\colhead{I(H$\alpha$)\tablenotemark{c}} \\
\colhead{} &
\colhead{[$10^{14}$\,cm$^{-2}$]} &
\colhead{[$10^{14}$\,cm$^{-2}$]} &
\colhead{[$10^{14}$\,cm$^{-2}$]} &
\colhead{[$10^{14}$\,cm$^{-2}$]} &
\colhead{[$10^{19}$\,cm$^{-2}$]} &
\colhead{} &
\colhead{} &
\colhead{} &
\colhead{} &
\colhead{[mR]}
}
\startdata
Mrk~290                                        &
	       1.70$\pm$0.14\tablenotemark{d}  &
	                                       &
	    $>$1.0$\;$                         &
	                                       &
               11.5$\pm$0.5                    &
            $-$1.10$\pm$0.06                   &
                                               &
         $>$$-$1.61$\pm$0.03                   &
                                               &
 	        187$\pm$10                     \\
Mrk~817                                        &
	       1.89$\pm$0.13\tablenotemark{e}  &
	                                       &
	                                       &
	                                       &
                3.0$\pm$0.1                    &
            $-$0.48$\pm$0.06                   &
                                               &
                                               &
                                               &
	                                       \\
Mrk~279                                        &
	       2.48$\pm$0.68\tablenotemark{f}  &
	                                       &
	                                       &
	                                       &
                3.1$\pm$0.4                    &
            $-$0.36$\pm$0.18                   &
                                               &
                                               &
                                               &
	                                       \\
Mrk~501                                        &
	    $<$2.0\tablenotemark{g}$\qquad\quad\;\;\;$ &
	                                       &
	                                       &
	                                       &
                1.6$\pm$0.1                    &
         $<$$-$0.16$\pm$0.06$\;\;\;$           &
                                               &
                                               &
                                               &
	                                       \\
Mrk~876                                        &
	                                       &
	       1.01$\pm$0.13\tablenotemark{h}  &
	                                       &
	    $>$0.4$\;$                         &
                2.3$\pm$0.2                    &
                                               &
            $-$1.41$\pm$0.08                   &
                                               &
         $>$$-$1.30$\pm$0.04                   &
	       $<$20                           \\
\enddata
\tablenotetext{a}{Quoted column densities 
reflect those derived using the $\tau_v$ technique of
Sembach \& Savage (1992).  Quoted uncertainties correspond
to the total statistical noise (their equation A27); continuum placement
uncertainties are not included here.}
\tablenotetext{b}{Based on Effelsberg H~I from Wakker et~al.(2001).}
\tablenotetext{c}{Based on observations made with the Wisconsin
H$\alpha$ Mapper: Mrk~290 (Wakker et~al. 1999a,b); Mrk~876 (Murphy et~al.
2000).}
\tablenotetext{d}{Based on the weighted mean of N(S~II\,$\lambda$1253) and
N(S~II\,$\lambda$1259).}
\tablenotetext{e}{Based on the weighted mean of N(S~II\,$\lambda$1250) and
N(S~II\,$\lambda$1253).}
\tablenotetext{f}{Based on uncertain N(S~II\,$\lambda$1250).}
\tablenotetext{g}{Based on N(S~II\,$\lambda$1253).  Highly uncertain upper
limit, due to poorly constrained local continuum.}
\tablenotetext{h}{Based on N(N~I\,$\lambda$1199).
Consistent with N(N~I\,$\lambda$1134.1)=
(1.6$\pm$0.6)$\times$10$^{14}$\,cm$^{-2}$ (Murphy et~al. 2000).}
\end{deluxetable}

\clearpage

\setlength{\topmargin}{0.0mm}
\epsscale{1.0}
\plotone{Gibson.fig1.eps}

\clearpage

\epsscale{1.0}
\plotone{Gibson.fig2.eps}

\clearpage

\epsscale{1.0}
\plotone{Gibson.fig3.eps}

\clearpage

\epsscale{1.0}
\plotone{Gibson.fig4.eps}

\clearpage

\epsscale{1.0}
\plotone{Gibson.fig5.eps}

\clearpage

\epsscale{1.0}
\plotone{Gibson.fig6.eps}

\clearpage

\epsscale{1.0}
\plotone{Gibson.fig7.eps}

\clearpage

\epsscale{1.0}
\plotone{Gibson.fig8.eps}

\clearpage

\epsscale{1.0}
\plotone{Gibson.fig9.eps}

\clearpage

\epsscale{1.0}
\plotone{Gibson.fig10.eps}

\clearpage

\epsscale{1.0}
\plotone{Gibson.fig11.eps}

\clearpage

\epsscale{1.0}
\plotone{Gibson.fig12.eps}

\end{document}